\documentclass[12pt,a4paper]{article}
\usepackage[latin1]{inputenc}
\usepackage{a4wide}
\usepackage{amsfonts}
\usepackage{amssymb}
\usepackage{amsmath}
\usepackage{bbm}
\usepackage{ifpdf}
\ifpdf
\usepackage[pdftex]{graphicx}
\usepackage[pdftex,unicode,implicit]{hyperref}
\hypersetup{%
  pdftitle    = {Stringy cosmic strings in matter coupled N=2, d=4 supergravity},
  pdfkeywords = {Supersymmetry, supergravity, strings, cosmic strings,
special geometry, BPS solutions},
  pdfauthor   = {Eric Bergshoeff, Jelle Hartong, Mechthild Hbscher and Toms Ortn},
  pdfcreator  = {pdf\LaTeXe\ with package \flqq hyperref\frqq},
  pdfproducer = {pdf\LaTeXe\ with package \flqq hyperref\frqq},
  pdfpagemode = None,  
  pdffitwindow= true,  
  unicode     = true,
  plainpages  = true,
  colorlinks  = true,  
  citecolor   = blue,
  urlcolor    = red,
  linkcolor   = black
}
\newcommand{\hepth}[1]{arXiv:{\tt
\href{http://www.arXiv.org/abs/hep-th/#1}{hep-th/#1}}}

\newcommand{\arxiv}[1]{{\tt
\href{http://www.arXiv.org/abs/#1}{arXiv:#1}}}
\else
  \usepackage[dvips]{graphicx}
  \usepackage[unicode,implicit]{hyperref}
  \newcommand{\hepth}[1]{arXiv:{\tt hep-th/#1}}

  \newcommand{\arxiv}[1]{{\tt arXiv:#1}}
\fi
\makeatletter
\@addtoreset{equation}{section}
\makeatother

\makeindex
\begin{document}

\begin{flushright}
\small
UG-07-07\\
IFT-UAM/CSIC-07-23\\
November $6^{\rm th}$ 2007\\
\normalsize
\end{flushright}

\begin{center}

\vspace{.7cm}

{\LARGE {\bf Stringy cosmic strings  \\[.5cm]
in  \\[.5cm]
matter coupled $N=2$, $d=4$ supergravity}}

\vspace{8mm}

\begin{center}
    Eric A. Bergshoeff$^{\dagger}$, Jelle Hartong$^{\dagger}$,
    Mechthild H\"{u}bscher$^{\ddagger}$
    and Tom\'{a}s Ort\'{\i}n$^{\ddagger}$  \\[5mm]
    {\small\slshape
    $\dagger$ Centre for Theoretical Physics, University of Groningen, \\
    Nijenborgh 4,
    9747 AG Groningen, The Netherlands \\
    {\upshape\ttfamily E.A.Bergshoeff, J.Hartong@rug.nl} \\[2mm]
    $\ddagger$ Instituto de F\'{i}sica Te\'{o}rica UAM/CSIC, \\ Facultad de
    Ciencias C-XVI,\\ C.U. Cantoblanco,\\ E-28049 Madrid, Spain \\
    {\upshape\ttfamily Mechthild.Huebscher@uam.es, Tomas.Ortin@cern.ch}}
\end{center}

\vspace{8mm}

{\bf Abstract}

\begin{quotation}
We extend the system of ungauged $N=2,d=4$ supergravity coupled to
vector multiplets and hypermultiplets with 2-form potentials. The
maximal number of 2-form potentials that one may introduce is
equal to the number of isometries of either the special K\"{a}hler
or quaternionic K\"{a}hler sigma model. We show that the local
supersymmetry algebra can be realized on the 2-form potentials.
These 2-forms couple electrically to strings which we refer to as
stringy cosmic strings. The 1/2 BPS bosonic world-sheet actions
for these strings are constructed and we discuss the properties of
the 1/2 BPS stringy cosmic string solutions.
\end{quotation}

\end{center}

\newpage
\pagestyle{plain}

\tableofcontents

\newpage

\section{Introduction}

When constructing a matter-coupled supergravity theory one usually
concentrates on the fields that describe the physical states of
the theory in question. Generically the bosonic states are
represented by the graviton, and a set of matter fields that
generically are differential forms of low rank $(d-2)/2\geq p\ge0$
for $d$ even and $(d-3)/2\geq p\ge0$ for $d$ odd, respectively. To
describe the coupling to branes one is naturally led to consider
the dual $(d-p-2)$-form potentials as well. For $p\neq 0$ and at
leading order, the construction of the dual potentials is rather
straightforward since the original low-rank differential form
fields always occur via their curvatures. This means that one may
even eliminate the potentials of the theory in favor of their
duals. However, at higher orders, there may be non-derivative
couplings and, while the dualization would still be possible, the
elimination would not. A prime example of this is the trilinear
coupling of the 3-form potential of $d=11$ supergravity.  In this
case one can introduce a dual 6-form potential without being able
to eliminate the 3-form potential.  This is related to the fact
that the 6-form field transforms under the gauge transformations
of the 3-form potential leading to a non-trivial bosonic gauge
algebra \cite{Cremmer:1998px}.

The situation is more involved for the scalar fields, i.e.~$p=0$
since often they appear via non-linear non-derivative couplings.
It is instructive to consider the explicit example of IIB
supergravity which has two scalars: the dilaton and the RR axion.
Together they parameterize the scalar coset
$SL(2,\mathbb{R})/U(1)$. The dualization of the RR axion is
straightforward since at leading order it only appears under a
derivative. The dual RR 8-form potential couples to the D7-brane.
However, the definition of the axion is basis-dependent. Using
another coordinate system for the $SL(2,\mathbb{R})/U(1)$ coset
manifold one can define a new axion $\chi^{\prime}$ which is
different from the RR axion as explained in
\cite{Bergshoeff:2007aa}.  Dualizing $\chi^{\prime}$, which is a
function of the old dilaton and RR axion, leads to a new 8-form
potential that is not related to the RR 8-form potential by any
$SL(2,\mathbb{R})$ duality transformation. To obtain a manifestly
$SL(2,\mathbb{R})$-covariant dualization prescription of all
possible axions one must dualize the Noether currents associated
to the presence of isometries of the scalar manifold. After all,
in an appropriate coordinate system, these isometries become shift
symmetries of given scalar fields. In the case of
$SL(2,\mathbb{R})/U(1)$ there are three isometries and this
procedure leads to three dual 8-form potentials.  Since there are
only two scalars and one cannot have more dual 8-form potentials
than scalars one finds that the triplet of 8-form potentials
satisfies a single duality-invariant constraint
\cite{Cremmer:1998px,Meessen:1998qm,Dall'Agata:1998va}. Another
way to see this is by noting that one of the three scalars on
which the isometries act as shifts does not correspond to a
(discrete) isometry of the quantum moduli space
$SL(2,\mathbb{Z})\backslash SL(2,\mathbb{R})/SO(2)$ so that
effectively only two 8-forms need to be considered.

The 8-form potentials of IIB supergravity play an important role
when discussing the supersymmetry properties of 7-branes in ten
dimensions \cite{Bergshoeff:2007aa,Bergshoeff:2006jj}. Likewise in
four dimensions 2-form potentials are dual to those scalars which
parameterize the Noether currents. They couple electrically to
1-dimensional branes which we refer to as stringy cosmic strings
in analogy with the terminology used in \cite{Greene:1989ya} where
a subset of the stringy cosmic strings of the
$SL(2,\mathbb{R})/U(1)$ coset was studied.

In this paper we generalize the case of the $SL(2,\mathbb{R})/U(1)$ coset in
four dimensions to $N=2$ supergravity coupled to an arbitrary number of vector and
hypermultiplets whereby we assume that the scalar sigma models admit some
isometry group. This is in no way a restrictive condition because without
isometries one cannot even define a 2-form potential. It was shown in
\cite{Freedman:1980us} that one cannot in general dualize just any scalar into
a 2-form potential. The objects to dualize are those Noether currents
associated with the isometries of the scalar sigma models which extend to be
symmetries of the full theory. Dualizing the Noether currents one obtains as
many 2-forms as there are isometries. In general the field strengths of these
2-forms satisfy constraints such that the number of 2-form degrees of freedom
equals the number of scalar degrees of freedom which occur in the Noether
currents.

We explicitly construct the Noether currents for all the duality symmetries of
ungauged $N=2, d=4$ supergravity coupled to both vector multiplets
and hypermultiplets. Via a
straightforward dualizing prescription we construct the 2-form potentials and
prove that the supersymmetry algebra can be closed on them. Once we have found
the explicit supersymmetry transformations for the 2-forms we proceed to
construct the leading terms of a half-supersymmetric world-sheet effective
action. Finally we discuss to some detail the properties of the
half-supersymmetric stringy cosmic string solutions.  The above program is first performed for the duality
symmetries associated with the scalars coming from the vector
multiplets and then repeated for the duality symmetries associated
with the scalars coming from the hypermultiplets.

In dualizing the 2-forms which are dual to the scalars of the vector multiplets
it turns out to be necessary to incorporate into the discussion both the
1-forms and their duals. This is because the gauge transformations of the
2-forms involve both the 1-forms and their duals. We will therefore also
briefly discuss the supersymmetry properties of the dual 1-forms and as a side
result construct world-line effective actions for 0-branes carrying an
arbitrary number of electric and magnetic charges. These 0-brane effective
actions may be used as sources for extreme supersymmetric black holes with
electric and magnetic charges.

This paper is organized as follows. In Section~\ref{sec-N2d4sugra} we give a
brief description of $N=2,d=4$ supergravity coupled to vector multiplets and hypermultiplets.
In Section~\ref{sec-1forms} we study dual 1-forms and their supersymmetry
transformation rules. These are used in
Section~\ref{sec-0braneeffectiveactions} to construct symplectic-invariant
0-brane word-line actions.  The symplectic invariance refers to the fact
that the world-line actions contain both the 1-forms and their duals.
In Section~\ref{sec-2forms} we construct the
2-forms dual to the scalars of the vector multiplets in three steps. In
Section~\ref{sec-noether} we construct the Noether current 1-forms associated
to the isometries of the special K\"{a}hler manifold.  They are on-shell
dualized into 2-forms in Section~\ref{sec-2formsdef}. The supersymmetry
transformations of these 2-forms are constructed in
Section~\ref{sec-2formssusy}. In Section~\ref{sec-1braneeffectiveactions} we
will apply our results to construct the stringy cosmic string world-sheet
effective actions. The supersymmetric stringy cosmic string solutions
associated to these effective actions are discussed in
Section~\ref{sec-stringsolutions}. In Sections
\ref{sec-hyper2forms} to \ref{hypersolutions} we repeat this
program for the isometries of the quaternionic K\"{a}hler manifold
which lead to the 2-forms dual to the hyperscalars. Our conclusions are contained in
Section~\ref{sec-conclusions}.


\section{Matter-coupled, ungauged, $N=2$, $d=4$ supergravity}
\label{sec-N2d4sugra}

Our starting point is $N=2,d=4$ ungauged supergravity coupled to $n_{V}$
vector and $n_{H}$ hypermultiplets. This is the same theory that was studied
in \cite{Meessen:2006tu}, whose conventions we use here\footnote{They are
  those of Ref.~\cite{Andrianopoli:1996cm} with some minor changes introduced
  in Refs.~\cite{Meessen:2006tu,Bellorin:2005zc}.}. In this
Section we will briefly review it for the sake of self-consistency, referring
the reader to \cite{Meessen:2006tu,Bellorin:2005zc}, the
reviews \cite{Andrianopoli:1996cm,kn:toinereview} and the original papers
\cite{deWit:1984pk,deWit:1984px} for more details. Our conventions have been
summarized in Appendix \ref{gammaspinor}.

The bosonic fields of the theory are those of the $N=2$, $d=4$ supergravity
multiplet (metric and graviphoton) and of $n_{V}$ vector multiplets ($n_V$
complex scalars and $n_V$ vectors) and $n_{H}$ hypermultiplets
($4n_H$ real scalars). The graviphoton together with the $n_V$
vectors are combined into the vector $A^{\Lambda}_{\mu}$ where
$\Lambda=0,1,\ldots,n_{V}$. The complex scalars will be denoted by $Z^i$ with
$i=1,\ldots,n_V$ while the real scalars will be denoted by $q^u$ with $u=1,\ldots,4n_H$.

The action of the bosonic fields of the theory is

\begin{equation}
\label{action}
\begin{array}{rcl}
 S & = & {\displaystyle\int} d^{4}x \sqrt{|g|}
\left[R +2\mathcal{G}_{ij^{*}}\partial_{\mu}Z^{i}
\partial^{\mu}Z^{*\, j^{*}} +2\mathsf{H}_{uv}\partial_{\mu}q^{u} \partial^{\mu}q^{v}
 \right. \\
& & \\
& & \left.
\hspace{2cm}
+2\Im{\rm m}\mathcal{N}_{\Lambda\Sigma}
F^{\Lambda\, \mu\nu}F^{\Sigma}{}_{\mu\nu}
-2\Re{\rm e}\mathcal{N}_{\Lambda\Sigma}
F^{\Lambda\, \mu\nu}{}^{\star}F^{\Sigma}{}_{\mu\nu}
\right]\, ,
\end{array}
\end{equation}

\noindent where the complex scalars $Z^i$ parameterize a special
K\"{a}hler manifold and where the real scalars $q^u$ parameterize
a quaternionic K\"{a}hler manifold. For their definitions and
properties we refer the reader to Appendices
\ref{sec-specialgeometry} and \ref{sec-QKG}. The metric on the
special K\"{a}hler manifold is denoted by $\mathcal{G}_{ij^{*}}$,
where the index $(j^*) i$ is a (anti-)holomorphic index. The field
strengths of the vectors $A^{\Lambda}_{\mu}$ are
$F^{\Lambda}_{\mu\nu}=\partial_{\mu}A^{\Lambda}_{\nu}-\partial_{\nu}A^{\Lambda}_{\mu}$.
The scalars couple to the vectors via the period matrix
$\mathcal{N}_{\Lambda\Sigma}$ whose definition is given in
Appendix \ref{sec-specialgeometry}. The last term in
\eqref{action} is topological with

\begin{equation}
{}^{\star}F^{\Sigma}{}_{\mu\nu}\equiv\tfrac{1}{2\sqrt{\vert
g\vert}}\,\epsilon_{\mu\nu\rho\sigma}F^{\Sigma\,\rho\sigma}\, .
\end{equation}

It is an important feature of the above action that the period matrix
$\mathcal{N}$ is only a function of the complex scalars $Z^{i}$ and $Z^{*\,i^*}$ of the vector
multiplets and does not depend on the quaternionic scalars $q^{u}$ of the
hypermultiplets. The vector and hypermultiplets only interact gravitationally.

The field strengths $F^{\Lambda}{}_{\mu\nu}$ of the vector potentials
$A^{\Lambda}{}_{\mu}$ satisfy the Bianchi identity

\begin{equation}
\label{BianchiA}
\nabla_{\nu}({}^{\star}F^{\Lambda})^{\nu\mu}=0\qquad\text{or}\qquad
dF^{\Lambda}=0\, ,
\end{equation}

\noindent
and the equation of motion

\begin{equation}\label{eomA}
\frac{1}{8\sqrt{\vert g\vert}}\frac{\delta S}{\delta A^{\Lambda}_{\mu}}=
\nabla_{\nu}({}^{\star}F_{\Lambda})^{\nu\mu}=0\, ,
\end{equation}

\noindent
where we have defined the dual vector field strength $F_{\Lambda}$
by

\begin{equation}
\label{eq:defdualfieldstrengths}
F_{\Lambda\, \mu\nu} \equiv
-\frac{1}{4\sqrt{|g|}}\frac{\delta S}{\delta {}^{\star}F^{\Lambda}{}_{\mu\nu}}
= \Re {\rm e}\mathcal{N}_{\Lambda\Sigma}F^{\Sigma}{}_{\mu\nu}
+\Im {\rm m}\mathcal{N}_{\Lambda\Sigma}{}^{*}F^{\Sigma}{}_{\mu\nu}\, .
\end{equation}

\noindent
The equation of motion \eqref{eomA} can be interpreted as a Bianchi identity
for the dual field strength $F_{\Lambda}$,

\begin{equation}\label{BianchidualA}
    dF_{\Lambda}=0\, ,
\end{equation}

\noindent
implying the local existence of $n_V+1$ dual vector fields $A_{\Lambda}$,
i.e.~locally $F_{\Lambda}=dA_{\Lambda}$. The equation of motion and Bianchi
identity for $A^{\Lambda}$, Eqs.~\eqref{eomA} and \eqref{BianchiA},
respectively, can be summarized as

\begin{equation}
\label{eq:symplecticBianchis}
d\mathcal{F}=0\, ,
\end{equation}

\noindent
where $\mathcal{F}$ is the $(2n_V+2)$-dimensional vector of field strengths

\begin{equation}\label{BianchiAanddualA}
\mathcal{F}\equiv
\left(
\begin{array}{c}
F^{\Lambda}\\
F_{\Lambda}\\
\end{array}
\right)\, .
\end{equation}

The Maxwell equations and Bianchi identities are left (formally) invariant by
the transformations of the vector field strengths

\begin{equation}\label{GLtrafo}
\mathcal{F}^{\prime}= \mathcal{S}\mathcal{F}\, ,\hspace{.5cm}
\mathcal{S}\equiv\left(
        \begin{array}{cc}
            A & B \\
            C & D \\
        \end{array}
    \right)\in GL(2n_V+2,\mathbb{R})\, ,
\end{equation}

\noindent
$A,B,C$ and $D$ being $(n_{V}+1)\times (n_{V}+1)$ matrices.  The
$(2n_V+2)$-dimensional vector of potentials

\begin{equation}
\label{eq:symplecticvector}
\mathcal{A}\equiv
\left(
\begin{array}{c}
A^{\Lambda}\\
A_{\Lambda}\\
\end{array}
\right)\, ,
\end{equation}

\noindent
whose local existence is implied by Eqs.~(\ref{eq:symplecticBianchis}),
transforms in the same way. However, since the dual potentials,
$A_{\Lambda}$, depend in a non-local way on the `fundamental' ones,
$A^{\Lambda}$, these transformations are non-local and are not symmetries of
the action, which only depends on the fundamental potentials, but
only of the Maxwell equations and Bianchi identities.

We have to take into account, however, that the definition of the dual field
strength $F_{\Lambda}$ involves the period matrix
$\mathcal{N}_{\Lambda\Sigma}$. In order to preserve this relation, the period
matrix must transform under the above $GL(2n_V+2,\mathbb{R})$ transformations
as

\begin{equation}\label{trafoperiodmatrix}
\mathcal{N}^{\prime}=(D\mathcal{N}+C)(B\mathcal{N}+A)^{-1}\, .
\end{equation}

\noindent
The period matrix $\mathcal{N}_{\Lambda\Sigma}$ is symmetric in its indices
$\Lambda$ and $\Sigma$. Demanding that this symmetry is preserved under the
transformation \eqref{trafoperiodmatrix} one finds that the matrices $A, B, C,
D$ must satisfy

\begin{equation}
D^{T}B=B^{T}D\,,\quad C^{T}A=A^{T}C
\quad\text{and}\quad
D^{T}A-B^{T}C=\mathbbm{1}\, ,
\end{equation}

\noindent
or

\begin{equation}\label{omega}
    \mathcal{S}^{T}\Omega \mathcal{S}=\Omega
    \hspace{.5cm}\text{with}\hspace{.5cm}
        \Omega \equiv\left(
        \begin{array}{lr}
            0 & -\mathbbm{1} \\
            \mathbbm{1} & 0 \\
        \end{array}
    \right)\, ,
\end{equation}

\noindent
so that $\mathcal{S}\in Sp(2n_V+2,\mathbb{R})$ and only this subgroup of
elements
$\mathcal{S}\in GL(2n_V+2,\mathbb{R})$ can be a symmetry of all the equations
of motion of the theory\footnote{This, in fact, is the largest possible
  electro-magnetic duality group of any Lagrangian depending on Abelian field
  strengths, scalars and derivatives of scalars as well as spinor fields
  \cite{Gaillard:1981rj}.}.

It can be checked that this condition is enough for the transformations to
leave invariant the Einstein equations as well, but, to be symmetries of all
the equations of motion, they have to leave invariant the scalar equations of
motion as well.

Since the period matrix is a function of the complex scalars,
$\mathcal{N}_{\Lambda\Sigma}=\mathcal{N}_{\Lambda\Sigma}(Z,Z^*)$, the
transformations \eqref{trafoperiodmatrix} induce transformations of the
complex scalars $Z^i$.  The kinetic term for $Z^i$ in \eqref{action} will be
invariant when the scalar transformations \eqref{trafoperiodmatrix} are
isometries of the metric $\mathcal{G}_{ij^{*}}$. Thus, out of the group
$GL(2n_V+2,\mathbb{R})$, only the subgroup $G_{V}$ of isometries of the
special K\"{a}hler manifold that can be embedded in $Sp(2n_V+2,\mathbb{R})$
is a symmetry of the full set of equations of motion and Bianchi identities.
In order for $G_{V}$ to be a symmetry of the complete supergravity theory, it must
satisfy some extra conditions that we will study in Section~\ref{sec-noether},
see \eqref{constraintQ}.
 There can be further
symmetries which are the isometries of the quaternionic K\"{a}hler
manifold, i.e.~isometries of the metric $\mathsf{H}_{uv}$. These
isometries are unrelated to the electromagnetic duality group
$Sp(2n_V+2,\mathbb{R})$. All these symmetries and the extended
objects associated to them will be the subject of this paper.

The fermionic fields of the theory are those of the $N=2$, $d=4$
supergravity multiplet (two gravitini $\Psi_{I\,\mu}\,,I=1,2$),
$n_V$ vector multiplets ($n_V$ gaugini $\lambda^{i\,I}$) and of
$n_H$ hypermultiplets ($2n_H$ hyperini
$\zeta_{\alpha}\,,\alpha=1,\ldots,2n_H$). We take all spinors to
be complex Weyl spinors. We define
$\lambda^{i^*}{}_{I}=(\lambda^{i\,I})^*$ and
$\zeta^{\alpha}=(\zeta_{\alpha})^*$. The index $\alpha$ is an
$Sp(2n_H)$ index where by $Sp(2n_H)$ we mean the compact
symplectic group $Sp(2n_H)\simeq U(4n_H)\cap Sp(4n_H,\mathbb{C})$.

The R-symmetry group of $N=2$, $d=4$ supergravity is $SU(2)\times U(1)$. The
$U(1)$ gauge connection is the K\"{a}hler connection 1-form, denoted by
$\mathcal{Q}$, and the spinors all carry a particular K\"{a}hler weight with
respect to $\mathcal{Q}$ (see Appendix \ref{sec-specialgeometry} for more
details).  The $SU(2)$
gauge connection is denoted by $\mathsf{A}_{I}{}^{J}$ and acts on
all objects which carry an $SU(2)$ index $I=1,2$ (see Appendix
\ref{sec-QKG} for more details about $\mathsf{A}_{I}{}^{J}$).

From this point on we will refer to the upper case Greek indices as symplectic
indices and to vectors $X$ given by

\begin{equation}
X=
\left(
\begin{array}{c}
X^{\Lambda}\\
X_{\Lambda}\\
\end{array}
\right)\,
\end{equation}

\noindent
as symplectic vectors. Given two symplectic vectors $X$ and $Y$ we define the
symplectic-invariant inner product, $\langle X\mid Y\rangle$, by

\begin{equation}
\langle\, X\mid Y\, \rangle=
X^{T} \Omega Y=X_{\Lambda}Y^{\Lambda}-X^{\Lambda}Y_{\Lambda}\, .
\end{equation}

When writing forms inside a symplectic inner product we will implicitly assume
that we are taking the exterior product of both. One should then keep in mind
that $\langle X_{(p)}\mid T Y_{(q)}\rangle=(-1)^{pq}\langle Y_{(q)}\mid T
X_{(p)}\rangle$, where $X_{(p)}$ and $Y_{(q)}$ are p- and q-forms,
respectively.

We next discuss the supersymmetry transformations (up to second order in
fermions) of all the fields of the theory. The supersymmetry transformations
of the bosonic fields are

\begin{align}
  \delta_{\epsilon} e^{a}{}_{\mu} \;& =\;
-{\textstyle\frac{i}{4}} \bar{\psi}_{I\, \mu}\gamma^{a}\epsilon^{I}
+\text{c.c.}\, ,
\label{eq:susytranse}\\
& \nonumber \\
  \delta_{\epsilon} A^{\Lambda}{}_{\mu} \;& =\;
{\textstyle\frac{1}{4}}
\mathcal{L}^{\Lambda}\,
\epsilon_{IJ}\bar{\psi}^{I}_{\mu}\epsilon^{J}
+{\textstyle\frac{i}{8}}\mathfrak{D}_i\mathcal{L}^{\Lambda}\,\epsilon_{IJ}
\bar{\lambda}^{Ii}\gamma_{\mu}
\epsilon^{J}
+\text{c.c.}\, ,
\label{eq:susytransAfundamental}\\
& \nonumber \\
  \delta_{\epsilon} Z^{i} \;& =\;
{\textstyle\frac{1}{4}} \bar{\lambda}^{Ii}\epsilon_{I}\, ,
\label{eq:susytransZ}
\\
& \nonumber \\
  \delta_{\epsilon} q^{u} \;& = \; \tfrac{1}{4}\mathsf{U}^{\alpha I\,u}\bar{\zeta}_{\alpha}\epsilon_{I}+\text{c.c.}\, ,
\label{eq:susytransq}
\end{align}

\noindent
where $\mathcal{L}^{\Lambda}$ is defined in Appendix \ref{sec-specialgeometry}
as the upper part of the symplectic section $\mathcal{V}$ in terms of which a
special K\"{a}hler manifold can be defined and where
$\mathfrak{D}_{i}\mathcal{L}^{\Lambda}$ is the K\"ahler-covariant derivative
of $\mathcal{L}^{\Lambda}$ on the special K\"{a}hler manifold.
The object $\mathsf{U}^{\alpha I\,u}$ which appears in
Eq.~\eqref{eq:susytransq} is the complex conjugate of the
so-called inverse Quadbein, i.e.~$\mathsf{U}^{\alpha
I\,u}=(\mathsf{U}_{\alpha I}{}^{u})^*$. A Quadbein, denoted by
$\mathsf{U}^{\alpha I}{}_{u}$, is a Vielbein of the quaternionic
K\"{a}hler manifold and is defined in Appendix \ref{sec-QKG}. The
index pair $\alpha I$ on a Quadbein originates from the fact that
the holonomy group of a quaternionic K\"{a}hler manifold is
$Sp(1)\times Sp(2n_H)$ with $Sp(1)\simeq SU(2)$. The index pair $\alpha
I$ is raised and lowered under complex conjugation, e.g.
$\mathsf{U}_{\alpha I\,u}=(\mathsf{U}^{\alpha I}{}_u)^*$.

The fermionic field supersymmetry transformations are

\begin{eqnarray}
\delta_{\epsilon}\psi_{I\, \mu} & = &
\mathfrak{D}_{\mu}\epsilon_{I}
+\epsilon_{IJ}T^{+}{}_{\mu\nu}\gamma^{\nu}\epsilon^{J}\, ,
\label{eq:gravisusyrule}\\
& & \nonumber \\
\delta_{\epsilon}\lambda^{iI} & = &
i\not\!\partial Z^{i}\epsilon^{I} +\epsilon^{IJ}\not\!G^{i\, +}\epsilon_{J}\, .
\label{eq:gaugesusyrule}\\
& & \nonumber \\
\delta_{\epsilon}\zeta_{\alpha} & = &
i\mathsf{U}_{\alpha I\,u}\not\!\partial q^{u}\epsilon^{I}\, ,
\label{eq:hypersusyrule}
\end{eqnarray}

\noindent
The derivative $\mathfrak{D}_{\mu}$ is the Lorentz, K\"ahler and $SU(2)$ covariant derivative acting on objects with nonzero K\"{a}hler weights and $SU(2)$ indices $I,J$. In particular, it acts on the local supersymmetry transformation parameter $\epsilon_{I}$ as

\begin{equation}
\mathfrak{D}_{\mu} \epsilon_{I} =
(\nabla_{\mu} \ +\ {\textstyle\frac{i}{2}}\ \mathcal{Q}_{\mu})\ \epsilon_{I}
\ +\ \mathsf{A}_{\mu\, I}{}^{J}\ \epsilon_{J}\, ,
\end{equation}

\noindent where $\mathcal{Q}_{\mu}$ is the pullback of the
K\"ahler connection defined in Eq.~\eqref{eq:K1form} and where
$\mathsf{A}_{\mu\, I}{}^{J}$ is the pull back of the $SU(2)$
connection $\mathsf{A}_{I}{}^{J}$ of the quaternionic-K\"ahler
manifold,

\begin{equation}
    \mathsf{A}_{\mu\, I}{}^{J}=\mathsf{A}_{u\, I}{}^{J}\partial_{\mu}q^u\, .
\end{equation}

\noindent In the variation of the gravitini the hyperscalars only
appear via the $SU(2)$ connection $\mathsf{A}_{\mu\, I}{}^{J}$,
while in the variation of the gaugini the hyperscalars do not
appear at all. The 2-forms $T^{+}$ and $G^{i\, +}$ appearing in Eqs.~\eqref{eq:gravisusyrule}
and \eqref{eq:gaugesusyrule} are the self-dual parts of the graviphoton and
matter vector field strengths, respectively. They can be written in a
manifestly symplectic-invariant form as

\begin{align}
T^{+} \;& =\;\langle\,  \mathcal{V}\mid \mathcal{F}\, \rangle\, ,\\
& \nonumber \\
G^{i\,+} \;& =\;
\tfrac{i}{2}\mathcal{G}^{ij^*}
\langle\, \mathfrak{D}_{j^*}\mathcal{V}^{*}\mid \mathcal{F}\, \rangle\, .
\end{align}

The commutator of two supersymmetry transformations on any of the fields
presented in this Section has the universal form

\begin{equation}
\label{eq:universalalgebra}
[\delta_{\eta},\delta_{\epsilon}]
=\delta_{\text{g.c.t.}}(\xi)+\delta_{\text{gauge}}(\Lambda)\, ,
\end{equation}

\noindent
where $\delta_{\text{g.c.t.}}(\xi)$ is an infinitesimal general coordinate
transformation with parameter $\xi^{\mu}$ and $\delta_{\text{gauge}}(\Lambda)$
is a $U(1)$ gauge transformation with parameter $\Lambda^{\Lambda}$. The
parameters $\xi^{\rho}$ and $\Lambda^{\Lambda}$ are given by the spinor
bilinears

\begin{align}
\label{gctparameter}
\xi^{\mu}
\;& \equiv \;
-\tfrac{i}{4}\bar{\eta}^{I}\gamma^{\mu}\epsilon_{I}
+\text{c.c.}\, , \\
& \nonumber \\
\label{oneformgaugeparameter0}
\Lambda^{\Lambda}
\; & \equiv \;
-\xi^{\rho}A^{\Lambda}_{\rho}
+\tfrac{1}{4}\left(\mathcal{L}^{\Lambda}\epsilon_{IJ}\bar\eta^{I}\epsilon^{J}
+\text{c.c.}\right)\, .
\end{align}

In the next Sections we will define new dual fields of $N=2,d=4$
supergravity which will satisfy the same universal algebra with the possible
addition of specific gauge transformations which do not act on the original
`fundamental' fields that we have introduced in this Section.


\section{The 1-forms}
\label{sec-1forms}

The $N=2,d=4$ supergravity theory coupled to $n_V$ vector multiplets contains
$n_V+1$ `fundamental' vector fields $A^{\Lambda}{}_{\mu}$ whose supersymmetry
transformation rules are given in Eq.~(\ref{eq:susytransAfundamental}). The
potentials $A^{\Lambda}{}_{\mu}$ couple electrically to charged particles. In
the next Section we will construct the leading terms of the bosonic part of
the $\kappa$-symmetric world-line effective actions for particles electrically
charged under $A^{\Lambda}{}_{\mu}$.

As we mentioned in Section~\ref{sec-N2d4sugra}, the equations of motion of the
potentials $A^{\Lambda}{}_{\mu}$, Eqs.~\eqref{eomA}, can be understood as
providing the Bianchi identities for a set of dual field strengths
$F_{\Lambda}$ defined in Eq.~(\ref{eq:defdualfieldstrengths}). These equations
imply the on-shell local existence of $n_V+1$ dual potentials $A_{\Lambda\,
  \mu}$. The dual potentials $A_{\Lambda\, \mu}$ couple electrically to
particles which are magnetically charged under the fundamental vector fields
$A^{\Lambda}{}_{\mu}$. In this Section we will derive the supersymmetry
transformation rules for the dual potentials $A_{\Lambda\, \mu}$. This result
will then be used in the next Section to construct the leading terms of the
bosonic part of the $\kappa$-symmetric world-line effective actions for
particles electrically charged under the $A_{\Lambda\, \mu}$.

The fundamental potentials and their duals can be seen as, respectively, the
upper and lower components of the symplectic vector $\mathcal{A}_{\mu}$
defined in Eq.~(\ref{eq:symplecticvector}). Electric-magnetic duality
transformations act linearly on it. This suggests the following Ansatz for the
supersymmetry transformation rule of $\mathcal{A}$:

\begin{equation}
  \delta_{\epsilon} \mathcal{A}_{\mu} \; =\;
{\textstyle\frac{1}{4}}
\mathcal{V}\,
\epsilon_{IJ}\bar{\psi}^{I}_{\mu}\epsilon^{J}
+{\textstyle\frac{i}{8}}\mathfrak{D}_{i}\mathcal{V}\,\epsilon_{IJ}
\bar{\lambda}^{Ii}\gamma_{\mu}
\epsilon^{J}
+\text{c.c.}\, .
\label{eq:susytransA}
 \end{equation}

\noindent
This Ansatz agrees with the supersymmetry transformation rule of the
fundamental potentials $A^{\Lambda}{}_{\mu}$ as given in
Eq.~(\ref{eq:susytransAfundamental}) and with the fact that the
$A^{\Lambda}{}_{\mu}$ transform linearly under $Sp(2n_V+2,\mathbb{R})$.
Indeed, the supersymmetry algebra closes on the symplectic vector of 1-forms
$\mathcal{A}_{\mu}$ with the above supersymmetry transformation rule. We find
for the commutator of two supersymmetries acting on $\mathcal{A}_{\mu}$,

\begin{equation}
\label{oneformcommutator}
[\delta_{\eta},\delta_{\epsilon}] \mathcal{A}_{\mu}
=\delta_{\text{g.c.t.}}(\xi)\mathcal{A}_{\mu}+\delta_{\text{gauge}}(\Lambda)\mathcal{A}_{\mu}\, .
\end{equation}

\noindent
The general coordinate transformation of $\mathcal{A}_{\mu}$ is given by

\begin{equation}
\label{eq:1formgct}
\delta_{\text{g.c.t.}}(\xi)\mathcal{A}_{\mu}=\pounds_{\xi} \mathcal{A}_{\mu}=
\xi^{\nu} \partial_{\nu} \mathcal{A}_{\mu}
+(\partial_{\mu} \xi^{\nu})\mathcal{A}_{\nu}\, ,
\end{equation}

\noindent
with $\pounds_{\xi}$ denoting the Lie derivative and where the infinitesimal
parameter $\xi^{\rho}$ is given in Eq.~(\ref{gctparameter}). The gauge
transformation of $\mathcal{A}_{\mu}$ is given by

\begin{equation}
\label{eq:Agaugetransformation}
    \delta_{\text{gauge}}(\Lambda)\mathcal{A}_{\mu}=\partial_{\mu}\Lambda\,,
\end{equation}

\noindent
where the gauge transformation parameter $\Lambda$ is the symplectic-covariant
generalization of $\Lambda^{\Lambda}$ as given in
Eq.~(\ref{oneformgaugeparameter}) and is given by

\begin{equation}
\label{oneformgaugeparameter}
\Lambda
\equiv
-\xi^{\rho}\mathcal{A}_{\rho}
+\tfrac{1}{4}\left(\mathcal{V}\epsilon_{IJ}\bar\eta^{I}\epsilon^{J}
+\text{c.c.}\right)\, .
\end{equation}


\section{World-line actions for 0-branes}
\label{sec-0braneeffectiveactions}

In this Section we will construct the leading terms of the bosonic part of a
$\kappa$-invariant world-line effective action for 0-branes that couple to the
1-form potentials $A^{\Lambda}{}_{\mu}$ and $A_{\Lambda\, \mu}$. In doing so
we will take into account the symplectic structure of the theory. The actions
will be invariant under symplectic transformations provided we also transform
an appropriate set of the charges, in the spirit of
Ref.~\cite{Bergshoeff:2006gs}.

It is clear that the 0-branes of $N=2,d=4$ supergravity coupled to $n_V$
vector multiplets can carry both electric charges $q_{\Lambda}$ and magnetic
charges $p^{\Lambda}$ with respect to the fundamental potentials
$A^{\Lambda}{}_{\mu}$. The couplings of the magnetic 0-branes are, however,
better described as electric couplings to the dual potentials $A_{\Lambda\,
  \mu}$. A 0-brane with symplectic charge vector

\begin{equation}
q\equiv
\left(
  \begin{array}{c}
p^{\Lambda} \\
q_{\Lambda} \\
\end{array}
\right)\, .
\end{equation}

\noindent
will couple electrically to the potential $\mathcal{A}$. The only
symplectic-invariant coupling is $\langle q\mid \mathcal{A}\rangle$.
We thus propose the following Wess--Zumino term

\begin{equation}
\int d\tau\, \langle\, q \mid \mathcal{A}_{\mu}\, \rangle\,
\frac{dX^{\mu}}{d\tau}\, ,
\end{equation}

\noindent
where $\tau$ is the world-line parameter and $X^{\mu}$ the embedding
coordinate of the 0-brane. This Ansatz is clearly the only one satisfying the
requirements of symplectic invariance and gauge invariance.

The corresponding kinetic term in the 0-brane action is not much more
difficult to guess. Symplectic invariance requires that the charges
$q_{\Lambda}$ and $p^{\Lambda}$ appear in a symplectic invariant combination
with the scalars in the tension. The simplest combination is just the central
charge

\begin{equation}
\mathcal{Z}=\langle\, q \mid \mathcal{V}\, \rangle\, ,
\end{equation}

\noindent
whose asymptotic absolute value is known to give the mass of supersymmetric
black holes of these theories. Then, the world-line effective action takes the
form

\begin{equation}
\label{eq:worldlineaction}
S= \int d\tau \ |\mathcal Z|\
\sqrt{\frac{dX^{\mu}}{d\tau}\frac{dX^{\nu}}{d\tau} g_{\mu\nu}(X)}
+\int d\tau \langle\, q\mid \mathcal
A_{\mu}\,\rangle\frac{dX^{\mu}}{d\tau}\, .
\end{equation}

Using the supersymmetry transformations \eqref{eq:susytranse},
\eqref{eq:susytransZ} and \eqref{eq:susytransA} we find that the action
\eqref{eq:worldlineaction} preserves half of the supersymmetries with the
projector given by

\begin{equation}
\label{eq:constraint0} \epsilon_{I}+i\frac{\mathcal{Z}}{|\mathcal
Z|} \epsilon_{IJ} \frac{\gamma_\tau}{\sqrt{g_{\tau\tau}}}
\epsilon^{J}=0\, ,
\end{equation}

\noindent
where the subindex $\tau$ means contraction of a space-time index $\mu$ with
$dX^{\mu}/d\tau$. This is the same constraint that the Killing spinors of
supersymmetric $N=2,d=4$ black holes satisfy
\cite{Meessen:2006tu,Behrndt:1997ny,LopesCardoso:2000qm,Bellorin:2006xr}.  In
the static gauge, $\dot{X}^{\mu}=dX^{\mu}/d\tau=\delta^{\mu}{}_{t}$, assuming
a static metric, so that $\sqrt{g_{tt}}=e^{0}{}_{t}$ and denoting by
$e^{i\alpha}$ the phase of the central charge $\mathcal{Z}$, the above
projector takes the form

\begin{equation}
\label{eq:constraint0-1} \epsilon_{I}+ie^{i\alpha}
\epsilon_{IJ}\gamma_{0}\epsilon^{J}=0\, .
\end{equation}

\noindent
This equation is satisfied for spinors of the form

\begin{equation}
\epsilon_{I}=|X|^{1/2}e^{\frac{i}{2}\alpha}\epsilon_{I\,0}\, ,
\hspace{1cm}
\epsilon_{I\, 0}+i\epsilon_{IJ}\gamma_{0}\epsilon^{J\, 0}=0\, ,
\end{equation}

\noindent
in which the $\epsilon_{I\, 0}$ are constant spinors and with $|X|$ some real
function.


\section{The 2-forms: the vector case}
\label{sec-2forms}

In this Section we will construct the most general 2-forms associated to the
isometries of the special K\"{a}hler manifold one can introduce in $N=2,d=4$
supergravity coupled to $n_V$ vector multiplets
and $n_H$ hypermultiplets. The 2-forms associated to the isometries of the quaternionic K\"{a}hler manifold
will be discussed in Section \ref{sec-hyper2forms}. For the subset of commuting isometries a similar program has been performed in \cite{Claus:1997fk} where also actions for the
dualized scalars, which are part of so-called vector-tensor multiplets, are given.


\subsection{The Noether current}
\label{sec-noether}

As explained in Section \ref{sec-N2d4sugra} only the group $G_{V}$
of isometries of the special K\"{a}hler manifold which can be
embedded in $Sp(2n_V+2,\mathbb{R})$ are symmetries of the full set
of equations of motion and Bianchi identities. Despite the fact
that these duality transformations only leave invariant the
equations of motion together with the Bianchi identities, it is
possible to construct a conserved Noether current associated to
this invariance \cite{Gaillard:1981rj}. This is because under
variations of the scalars
$\delta_Z\mathcal{L}+\delta_{Z^{*}}\mathcal{L}$ the Lagrangian is
invariant up to the divergence of an anomalous current, denoted
here and in \cite{Gaillard:1981rj} by $\hat{J}^{\mu}$. Hence, we
have

\begin{equation}
\delta_Z\mathcal{L}+\delta_{Z^{*}}\mathcal{L}=
-\partial_{\mu}(\sqrt{\vert g\vert}\hat J^{\mu})\, .
\end{equation}
In the case of $p$-brane actions coupled to supergravity the Noether current
associated to the super-Poincar\'{e} invariance of the coupled system contains
a similar anomalous contribution \cite{deAzcarraga:1989gm}, which is known to
give rise to central charges in the supersymmetry algebra.

Applying the Noether theorem we get

\begin{equation}
\partial_{\mu}
\left(
\delta Z^{i}
\frac{\partial\mathcal{L}}{\partial(\partial_{\mu}Z^{i})}
+\delta Z^{*i^{*}}
\frac{\partial\mathcal{L}}{\partial(\partial_{\mu}Z^{*i^{*}})}
\right)
=
-\partial_{\mu}(\sqrt{\vert g\vert}\hat J^{\mu})\, ,
\end{equation}

\noindent
so that the Noether current

\begin{equation}
J_{N}^{\mu}=
\delta Z^{i}\frac{1}{\sqrt{\vert g\vert}}
\frac{\partial\mathcal{L}}{\partial(\partial_{\mu}Z^{i})}
+\delta Z^{*i^{*}}\frac{1}{\sqrt{\vert g\vert}}
\frac{\partial\mathcal{L}}{\partial(\partial_{\mu}Z^{*i^{*}})}
+\hat J^{\mu}\, ,
\end{equation}

\noindent
is covariantly conserved, i.e.~$\nabla_{\mu}J_{N}^{\mu}=0$. In this Subsection
we will compute $J_{N}^{\mu}$ for the isometries of the K\"ahler metric
$\mathcal{G}_{ij^{*}}$ which are embedded in $Sp(2n_V+2,\mathbb{R})$.

Infinitesimally, the symmetries under consideration act on the complex scalars
as

\begin{equation}\label{deltaZ}
\delta Z^{i} = \alpha^{A}k_{A}{}^{i}(Z)\, ,
\end{equation}

\noindent
where the $k_{A}{}^{i}(Z)$ are ${\rm dim}\, G_{V}$ holomorphic Killing
vectors\footnote{The holomorphicity of the components $k_{A}{}^{i}$ follows
  from the Killing equation.} ($A=1,\cdots,{\rm dim}\, G_{V}$) and where
$\alpha^{A}$ denotes a set of real infinitesimal parameters. The Lie brackets
of the Killing vectors give the Lie algebra of $G_{V}$ with structure
constants $f_{AB}{}^{C}$,

\begin{equation}
\label{structurecst}
[k_{A},k_{B}]=-f_{AB}{}^{C}k_{C}\, ,
\end{equation}

\noindent
where $k_{A}=k_{A}{}^{i} \partial_{i}+k_{A}{}^{*\,i^{*}} \partial_{i^{*}}$.

On the vector field strengths the symmetries act as an infinitesimal
$Sp(2n_V+2,\mathbb{R})$ transformation

\begin{equation}
\label{infinitrafo}
\delta\mathcal{F}= T\mathcal{F}\, ,
\end{equation}

\noindent where $T\in \mathfrak{sp}(2n_{V}+2,\mathbb{R})$,
i.e.~$T^{T}\Omega+\Omega T=0$. The matrix $T$ can be expressed as
a linear combination of the generators of the isometry group $G_V$
of $\mathcal{G}_{ij^{*}}$ that is embedded in
$\mathfrak{sp}(2n_V+2,\mathbb{R})$. In other words,

\begin{equation}\label{algebraT}
T=\alpha^{A}T_{A}\, ,
\hspace{1cm}
[T_{A},T_{B}]=f_{AB}{}^{C}T_{C}\, ,
\hspace{1cm}
T_{A}\in \mathfrak{sp}(2n_V+2,\mathbb{R})\, .
\end{equation}

\noindent
On the other hand, if

\begin{equation}
\label{chargematrixQ}
T=
\left(\begin{array}{cc}
a & b \\
c & d \\
\end{array}
\right)\, ,
\end{equation}

\noindent
then, the condition $T^{T}\Omega+\Omega T=0$ implies

\begin{equation}
\label{parameters}
c^{T}=c\,,\quad b^{T}=b\,,\quad\text{and}\quad a^{T}=-d\, .
\end{equation}

To find the current $\hat{J}^{\mu}$ we start by writing the Lagrangian of
\eqref{action} in the following form

\begin{equation}
\mathcal{L}= {\textstyle\frac{1}{2}}
F^{\Lambda}{}_{\mu\nu}
\frac{\partial\mathcal{L}}{\partial F^{\Lambda}{}_{\mu\nu}}+
{\mathcal{L}}_{\text{inv}}\,,
\end{equation}

\noindent
where

\begin{equation}
\mathcal{L}_{\text{inv}}=\sqrt{\vert g\vert}
\left[R+2\mathcal{G}_{ij^{*}}
\partial_{\mu}Z^{i}\partial^{\mu}Z^{*j^{*}}\right]\, ,
\end{equation}

\noindent
is the part of the Lagrangian that is invariant under \eqref{deltaZ}
and where

\begin{equation}
\frac{\partial\mathcal{L}}{\partial F^{\Lambda}{}_{\mu\nu}}=
-4\sqrt{\vert g\vert}\star F_{\Lambda}{}^{\mu\nu}\,.
\end{equation}

\noindent
Next we compute the variation of $\mathcal{L}$ with respect to the variation of
the scalars

\begin{equation}
\delta_Z\mathcal{L}+\delta_{Z^{*}}\mathcal{L}=
\delta\mathcal{L}-\delta_F\mathcal{L}\,,
\end{equation}

\noindent
where $\delta\mathcal{L}$ is the total variation and $\delta_F\mathcal{L}$
denotes the variation of $\mathcal{L}$ with respect to the field strength
$F^{\Lambda}_{\mu\nu}$. The total variation of $\mathcal{L}$ under the
transformations \eqref{deltaZ} and \eqref{infinitrafo} is

\begin{align}
\delta\mathcal{L} = &
\delta\left(-2\sqrt{\vert g\vert}
F^{\Lambda}{}_{\mu\nu}\star F_{\Lambda}{}^{\mu\nu}\right)=
-2\sqrt{\vert g\vert}
\left[\star F_{\Lambda}{}^{\mu\nu}b^{\Lambda\Sigma}
F_{\Sigma\, \mu\nu}
+\star F^{\Lambda\, \mu\nu}c_{\Lambda\Sigma}F^{\Sigma}{}_{\mu\nu}\right]\, ,
\end{align}

\noindent
where we have used Eqs.~\eqref{parameters}. The variation,
$\delta_F\mathcal{L}$, is

\begin{equation}
\delta_F\mathcal{L}= \delta F^{\Lambda}{}_{\mu\nu}
\frac{\partial\mathcal{L}}{\partial {F^{\Lambda}}_{\mu\nu}}=
-4\sqrt{\vert g\vert} \left[\star
F_{\Lambda}{}^{\mu\nu}a^{\Lambda}_{\;\;\Sigma}
F^{\Sigma}{}_{\mu\nu}+ \star
F_{\Lambda}{}^{\mu\nu}b^{\Lambda\Sigma}F_{\Sigma\, \mu\nu}\right]\,.
\end{equation}

\noindent
Using once again Eqs.~\eqref{parameters} it then follows that

\begin{equation}
\label{vartotL-varFL}
\delta\mathcal{L}-\delta_F\mathcal{L}=2\sqrt{\vert g\vert}
\langle\, \star\mathcal{F}^{\mu\nu} \mid T \mathcal{F}_{\mu\nu}\, \rangle\,.
\end{equation}

\noindent
The result Eq.~\eqref{vartotL-varFL} can be written as
the divergence of an anomalous current $\hat J$\, i.e.~one can
show, using Eqs.~\eqref{BianchiA} and \eqref{eomA}, that

\begin{equation}
-\partial_{\mu}(\sqrt{\vert g\vert}\hat{J}^{\mu})
=\delta\mathcal{L}-\delta_F\mathcal{L}\, ,
\end{equation}

\noindent
where $\hat{J}^{\mu}$ is given by

\begin{align}
\label{Jhat} \hat{J}^{\mu}= -4 \langle\,
\star\mathcal{F}^{\mu\nu} \mid T \mathcal{A}_{\nu}\, \rangle\, .
\end{align}

\noindent
At the same time we have for the right hand-side of this equation

\begin{equation}
\delta\mathcal{L}-\delta_F\mathcal{L}=
\delta_Z\mathcal{L}+\delta_{Z^{*}}\mathcal{L}=
\partial_{\mu}\left(\delta Z^{i}
\frac{\partial\mathcal{L}}{\partial(\partial_{\mu}Z^{i})}
+\delta Z^{*i^{*}}
\frac{\partial\mathcal{L}}{\partial(\partial_{\mu}Z^{*i^{*}})}\right)\, ,
\end{equation}

\noindent
so that the Noether current, $J_{N}^{\mu}$, is given by

\begin{equation}
\label{Noether}
J_{N}^{\mu}=
\delta Z^{i}\frac{1}{\sqrt{\vert g\vert}}
\frac{\partial\mathcal{L}}{\partial(\partial_{\mu}Z^{i})}+
\delta Z^{*i^{*}}\frac{1}{\sqrt{\vert g\vert}}
\frac{\partial\mathcal{L}}{\partial(\partial_{\mu}Z^{*i^{*}})}+
\hat{J}^{\mu}\,,
\end{equation}

\noindent
with $\hat{J}^{\mu}$ given by Eq.~\eqref{Jhat}, and satisfies

\begin{equation}
    \partial_{\mu}\left(\sqrt{\vert g\vert}J_N^{\mu}\right)=0\,.
\end{equation}

Under gauge transformations of the 1-form potentials $\mathcal{A}$
the anomalous current
$\hat{J}^{\mu}$ and hence $J_{N}^{\mu}$ are not invariant: they transform as
the divergence of an anti-symmetric tensor. We will have to take this point
into account in the next subsection
when dualizing the Noether current into a 2-form.

It will be convenient to write the scalar part of the Noether current,
i.e.~the part $J_N-\hat J$, in terms of the symplectic sections $\mathcal{V}$
instead of the physical scalars since $\mathcal{V}$ transforms linearly under
$Sp(2n_V+2,\mathbb{R})$.  This is achieved using

\begin{equation}\label{deltaV}
\delta\mathcal{V}=
\delta Z^{i}\partial_i\mathcal{V}
+\delta Z^{*i^{*}}\partial_{i^{*}}\mathcal{V}\, ,
\end{equation}

\noindent and Eqs.~\eqref{eq:SGDefFund} and \eqref{eq:SGProp1}. We
have

\begin{equation}
\delta Z^{i}\frac{\partial\mathcal{L}}{\partial(\partial_{\mu}Z^{i})}=
-2i\sqrt{\vert g\vert}
\langle\, \delta\mathcal{V}\mid\mathfrak{D}^{\mu}\mathcal{V}^{*}\,\rangle\, .
\end{equation}

\noindent
Hence, the Noether current \eqref{Noether} can be expressed in terms
of $\mathcal{V}$ as

\begin{equation}\label{Noethercurrent}
J_{N}^{\mu}=
-2i\langle\, \delta\mathcal{V}\mid\mathfrak{D}^{\mu}\mathcal{V}^{*}\, \rangle
+\mathrm{c.c.} +\hat{J}^{\mu}\, .
\end{equation}

We continue to find an explicit expression for $\delta\mathcal{V}$. The
symplectic sections transform under global $Sp(2n_V+2,\mathbb{R})$ and under
local K\"{a}hler transformations. The K\"{a}hler potential transforms as

\begin{equation}
\label{eq:Kpotentialtransformation}
\delta_{\alpha}\mathcal{K}\equiv
\pounds_{\alpha^{A}k_{A}}\mathcal{K}=
\alpha^{A}\left(k_{A}{}^{i}\partial_{i}\mathcal{K}
+k_{A}{}^{*\,i^{*}}\partial_{i^{*}}\mathcal{K}\right)=
\lambda(Z)+\lambda^{*}(Z^{*})\, ,
\hspace{1cm}
\lambda(Z)=\alpha^{A}\lambda_{A}(Z)\, .
\end{equation}

\noindent
It can be shown that the functions $\lambda_{A}(Z)$ satisfy

\begin{equation}
k_{A}^{i}\partial_{i}\lambda_{B}
-k_{B}^{i}\partial_{i}\lambda_{A} =-f_{AB}{}^{C}\lambda_{C}\, .
\end{equation}

\noindent
When $\lambda\neq 0$ all the objects of the theory with non-zero K\"ahler
weight (in particular all the spinors and the symplectic section
$\mathcal{V}$) will feel the effect of the symplectic transformation through a
K\"ahler transformation. Infinitesimally one has

\begin{equation}
\delta_{\text{K\"{a}hler}}\mathcal{V}=
-{\textstyle\frac{1}{2}}(\lambda-\lambda^{*})\mathcal{V}\, ,
\end{equation}

\noindent
as follows from Eq.~\eqref{KahlertrafoV}. Next we introduce the momentum map,
denoted by $\mathcal{P}^0_A$ and defined by

\begin{equation}
\mathcal{P}_A^0\equiv ik_{A}{}^{i}\partial_i\mathcal{K}-i\lambda_A\,.
\end{equation}

\noindent
One then readily shows that $\delta\mathcal{V}$, given via equations
\eqref{deltaV} and \eqref{deltaZ}, can be written as

\begin{equation}
\delta\mathcal{V}=
\alpha^A\left(k_{A}{}^{i}\mathcal{D}_i\mathcal{V}+i\mathcal{P}_A^0\mathcal{V}-
{\textstyle\frac{1}{2}}(\lambda_{A}-\lambda^{*}_{A})\mathcal{V}\right)\,.
\end{equation}

\noindent
Since $\mathcal{V}$ only transforms under symplectic and K\"{a}hler
transformations we conclude\footnote{Actually, this is a consequence of
  requiring that the reparametrizations generated by the Killing vectors
  preserve not just the metric but the whole special K\"ahler geometry. This
  is what we are implicitly doing here and it is a condition necessary to have
  symmetries of the complete supergravity theory and not just of the bosonic
  equations of motion. We thank Patrick Meessen for a useful discussion
 on this point.} that we must have

\begin{equation}
\label{deltaVII}
\delta\mathcal{V}=T\mathcal{V}-
{\textstyle\frac{1}{2}}(\lambda-\lambda^{*})\mathcal{V}\,,
\qquad\text{where}\qquad T\mathcal{V}=
\alpha^A\left(k_{A}{}^{i}\mathcal{D}_i\mathcal{V}
+i\mathcal{P}_A^0\mathcal{V}\right)\, ,
\end{equation}

\noindent where $T$ is a generator of $\mathfrak{sp}(2n_{V}+2)$.
Taking the product of the r.h.s.~of the second equation with
$\mathcal{V}$ we get the additional condition that the generators
of $G_{V}$ must satisfy:

\begin{equation}
\label{constraintQ}
\langle\, \mathcal{V}\mid T_A\mathcal{V}\, \rangle=0\, .
\end{equation}

\noindent
The set of generators $T_A$ which satisfy the constraint \eqref{constraintQ}
and which form a subgroup of $\mathfrak{sp}(2n_V+2,\mathbb{R})$ is sometimes
referred to as the \textit{duality symmetry Lie algebra} \cite{deWit:1995tf}.

\noindent
Since, on the other hand

\begin{equation}
\delta\mathcal{V}= \pounds_{\alpha^{A}k_{A}}\mathcal{V}=
\alpha^{A}\left(k_{A}{}^{i}\partial_{i}\mathcal{V}
+k_{A}{}^{*\,i^{*}}\partial_{i^{*}}\mathcal{V}\right)\, ,
\end{equation}

\noindent
we can write

\begin{equation}
\label{isometrycondition}
\pounds_{\alpha^{A}k_{A}}\mathcal{V}
-T\mathcal{V}+{\textstyle\frac{1}{2}}(\lambda-\lambda^{*})\mathcal{V}=0\, ,
\end{equation}

\noindent
as the necessary and sufficient condition for the transformation to be a
symmetry of the supergravity theory\footnote{This condition can be read in two
  different ways: the Lie derivative of the section $\mathcal{V}$ has to
  vanish up to symplectic and K\"ahler transformations or the symplectic- and
  K\"ahler-covariant Lie derivative of $\mathcal{V}$ has to vanish
  identically.}.

One verifies that the above way of writing the action of $T$ on
$\mathcal{V}$, see
Eq.~(\ref{deltaVII}), satisfies Eq.~\eqref{algebraT}. By decomposing
$T\mathcal{V}$ into the complete basis
$\{\mathcal{V},\mathfrak{D}_i\mathcal{V},\mathcal{V}^*,\mathfrak{D}_{i^*}\mathcal{V}^*\}$
for the space of symplectic sections (see Appendix \ref{sec-specialgeometry}
below Eq.~\eqref{eq:SGProp1}) we find

\begin{equation}
\label{momentummap}
\mathcal{P}_A^0=-\langle\mathcal{V}\mid T_A\mathcal{V}^*\rangle\,,
\hspace{.5cm}
\text{and}
\hspace{.5cm}
k_{A}{}^{i}=-i\mathcal{G}^{ij^*}\partial_{j^*}\mathcal{P}_A^0\,.
\end{equation}
Substituting \eqref{deltaVII} into expression \eqref{Noethercurrent} we obtain
a manifestly symplectic-invariant expression for the Noether current

\begin{equation}
J_{N\mu}= 2i\langle\, \mathfrak{D}_{\mu}\mathcal{V}^{*}\mid
T\mathcal{V}\, \rangle +\mathrm{c.c.} -4 \langle\,
\star\mathcal{F}_{\mu\nu} \mid T \mathcal{A}^{\nu}\, \rangle\, .
\end{equation}


\subsection{Dualizing the Noether current}
\label{sec-2formsdef}


In form notation the conservation of the Noether current 1-form $J_{N}$ is
just $d\star J_{N}=0$. We can define a 3-form\footnote{Of course, we have
  ${\rm dim}\, G_{V}$ Noether currents and as many dual 3-forms $G_{A}$ but it
  is convenient to work with $G=\alpha^{A}G_{A}$.} $G=\star J_{N}$, which
satisfies $dG=0$, so that locally $G=dB$. Note that $G$ is not
gauge invariant because $J_{N}$ is not, either, due to the term
$\hat{J}$ ($\delta_{\text{gauge}}G=\delta_{\text{gauge}}\hat{J}$).
We can write this term in the form

\begin{equation}
    \star\hat{J}=-4\langle\, \mathcal{F}\mid T \mathcal{A}\, \rangle\, ,
\end{equation}

\noindent
where the exterior product between the forms in the symplectic inner product
is always assumed and as a result the 2-form $B$ gauge transformation is given
by

\begin{equation}\label{gaugetrafoB}
\delta_{\text{gauge}} B=d\Lambda_{1}
-4\langle\, \mathcal{F}\mid T \Lambda\, \rangle\, ,
\end{equation}

\noindent
where the symplectic vector $\Lambda$ is defined through
Eq.~(\ref{eq:Agaugetransformation}).

We can define the following gauge-invariant 2-form field strength

\begin{equation}
\label{eq:H}
H=dB+4\langle\, \mathcal{F}\mid T \mathcal{A}\, \rangle\, .
\end{equation}

\noindent
It is then clear that $H$ is dual to the scalar part of the Noether current
$J_{N}$,

\begin{equation}
\label{dualityforH}
    H=\star(J_{N}-\hat{J})\, .
\end{equation}

\noindent
The scalar part of the Noether current is proportional to the Killing vectors.
At any given point there are only $2n_{V}$ (real) independent vectors. Thus,
if we allow for $Z^i$-dependent coefficients, in general we will find linear
combinations of scalar parts of the Noether currents. As a result, there will
be as many constraints on the 2-form field strengths $H_{A}$ and, at most
there will be $2n_{V}$ independent real 2-forms.


\subsection{The 2-form supersymmetry transformation}
\label{sec-2formssusy}

In the previous Subsection we have constructed a set of 2-forms associated to
the isometries of the special K\"{a}hler manifold of ungauged $N=2,d=4$
supergravity and we have found their gauge transformations. Our goal in this
Section is to find their supersymmetry transformations. The main requirement
that the proposed supersymmetry transformation of the 2-form $B$ must satisfy
is that the commutator agrees with the universal local supersymmetry algebra
of the theory given in Eq.~(\ref{eq:universalalgebra}) and which may be
extended to include 2-forms to

\begin{equation}
\label{commutatoralgebra}
[\delta_{\eta},\delta_{\epsilon}]=
\delta_{\text{g.c.t.}}(\xi)+\delta_{\text{gauge}}(\Lambda)
+\delta_{\text{gauge}}(\Lambda_{1})\, .
\end{equation}

\noindent
The expressions for $\xi$ and $\Lambda$ are given by Eqs.~\eqref{gctparameter}
and \eqref{oneformgaugeparameter}, respectively. The 2-form gauge
transformation parameter $\Lambda_{1}$ is to be found in terms of $\eta$ and
$\epsilon$.

Since $B$ is defined by $dB=\star J_{N}$, the commutator of two supersymmetry
variations on $B$ must close into the algebra \eqref{commutatoralgebra}. We
have

\begin{equation}
\label{gctA2}
    \delta_{\text{g.c.t.}}(\xi)B_{\mu\nu}=
\pounds_{\xi}B_{\mu\nu}=\xi^{\rho}\partial_{\rho}B_{\mu\nu}
+(\partial_{\mu}\xi^{\rho})B_{\rho\nu}
+(\partial_{\nu}\xi^{\rho})B_{\mu\rho}=\xi^{\rho}(dB)_{\rho\mu\nu}-
    2\partial_{[\mu}\left(\xi^{\rho}B_{\nu]\rho}\right)\,,
\end{equation}

\noindent
with $\pounds_{\xi}B_{\mu\nu}$ the Lie derivative of $B_{\mu\nu}$ with respect
to $\xi^{\rho}$. Further, $\delta_{\text{gauge}}(\Lambda_{1})B_{\mu\nu}$ is given in
Eq.~\eqref{gaugetrafoB}. Hence, the supersymmetry transformations of
$B_{\mu\nu}$ must lead to the commutator

\begin{equation}
\label{commutatorA2} [\delta_{\eta},\delta_{\epsilon}]B_{\mu\nu}
=\xi^{\rho}\tfrac{1}{\sqrt{\vert g\vert}}
\epsilon_{\rho\mu\nu\sigma}J_{N}{}^{\sigma} -4\langle\,
\mathcal{F}_{\mu\nu}\mid T \Lambda\, \rangle
+2\partial_{[\mu}\left(\Lambda_{\nu]}-\xi^{\rho}B_{\nu]\rho}\right)\,
,
\end{equation}

\noindent
where we have substituted the duality relation, Eq.~\eqref{dualityforH}, for
$(dB)_{\mu\rho\sigma}$ in \eqref{gctA2}.

We make the following Ansatz for the supersymmetry transformation
of $B_{\mu\nu}$ (up to second order in fermions),

\begin{eqnarray}
\label{ansatzsusyA2}
\delta_{\epsilon}B_{\mu\nu}
& = &
a\langle\, \mathfrak{D}_{i}\mathcal{V}\mid T\mathcal{V}^{*}\,\rangle\,
\bar\epsilon_{I}\gamma_{\mu\nu}\lambda^{iI}
+\text{c.c.}
\nonumber\\
& & \nonumber \\
& &
+b\langle\, \mathcal{V}\mid T\mathcal{V}^{*}\, \rangle\,
\bar\epsilon^{I}\gamma_{[\mu}\psi_{I\nu]}
+\text{c.c.}\nonumber\\
& & \nonumber \\
&&
+c\langle\, \mathcal{A}_{[\mu}\mid
T\delta_{\epsilon}\mathcal{A}_{\nu]}\, \rangle\, .
\end{eqnarray}

\noindent
This Ansatz is based on the requirement that all terms must have K\"{a}hler
weight zero and that the 2-forms are real valued. The matrix $T$ satisfies
Eq.~\eqref{constraintQ}.

We evaluate the commutator as follows. First we perform standard gamma matrix
manipulations, change the order of the spinors, evaluate the complex
conjugated terms and use relations from special geometry. Exhausting all such
operations using formulae from Appendices \ref{gammaspinor} and
\ref{sec-specialgeometry} leads to the following expression for the commutator

\begin{align}
\label{commutator1} [\delta_{\eta},\delta_{\epsilon}]B_{\mu\nu} &=
4ia\xi^{\sigma}\tfrac{1}{\sqrt{\vert
g\vert}}\epsilon_{\sigma\mu\nu\rho}\left[ \langle\,
\mathfrak{D}^{\rho}\mathcal{V}\mid T\mathcal{V}^{*}\, \rangle
-\langle\, \mathfrak{D}^{\rho}\mathcal{V}^{*}\mid T\mathcal{V}\,
\rangle\right]
\nonumber\\
& \nonumber \\
&
\left[
+4ia\langle\, \mathfrak{D}_{i}\mathcal{V}\mid T\mathcal{V}^{*}\,\rangle
\mathcal{G}^{ij^{*}}
\langle\, \mathfrak{D}_{j^{*}}\mathcal{V}^{*}\mid \mathcal{F}_{\mu\nu}\,\rangle
\epsilon^{IJ}\bar\eta_{I}\epsilon_{J}
\right.
\nonumber\\
& \nonumber \\
&
\left.
-2b\langle\,\mathcal{V}\mid T\mathcal{V}^{*}\,\rangle
\langle\, \mathcal{V}^{*}\mid \mathcal{F}_{\mu\nu}\,\rangle
\epsilon^{IJ}\bar{\eta}_{I}\epsilon_{J}
+\text{c.c.}
\right]
\nonumber\\
& \nonumber \\
&
-8a\xi_{[\nu}\partial_{\mu]}
\langle\, \mathcal{V}\mid T\mathcal{V}^{*}\, \rangle
+4ib\langle\, \mathcal{V}\mid T\mathcal{V}^{*}\, \rangle
\partial_{[\mu}\xi_{\nu]}+c\langle \mathcal{A}_{[\mu}\mid
[\delta_{\eta},\delta_{\epsilon}]\mathcal{A}_{\nu]}\, \rangle\,,
\end{align}

\noindent
where it has been assumed that $a$ and $ib$ are real parameters. The parameter
$\xi^{\rho}$ is given by \eqref{gctparameter}. The notation
$[\cdots+\text{c.c.}]$ means that one should take the complex conjugate of
whatever is written on the left within the brackets. The parameter $a$ has
been chosen to be real in order to obtain the scalar part of the Noether
current in the first line of \eqref{commutator1}. The parameter $ib$ has been
chosen to be real so that the K\"{a}hler connection 1-form $\mathcal{Q}_{\mu}$
appearing in $\delta_{\epsilon}\Psi_{I\,\mu}$ cancels when adding the complex
conjugated terms. We then take $2b=4ia$ so that the first and the second term
of the third line of Eq.~\eqref{commutator1} combine into a 2-form gauge
transformation parameter.  Expression \eqref{commutator1} is further
manipulated using the completeness relation Eq.~\eqref{completeness}. This is
the step where we impose the condition that $T$ must satisfy
Eq.~\eqref{constraintQ}. Using next the result for the 1-form commutator,
Eq.~\eqref{oneformcommutator}, to write out the term proportional to $c$ in
\eqref{commutator1}, we obtain

\begin{align}
[\delta_{\eta},\delta_{\epsilon}]B_{\mu\nu}
&=4ia\xi^{\sigma}\tfrac{1}{\sqrt{\vert
g\vert}}\epsilon_{\sigma\mu\nu\rho} \left[ \langle\,
\mathfrak{D}^{\rho}\mathcal{V}\mid T\mathcal{V}^{*}\,\rangle
-\langle\, \mathfrak{D}^{\rho}\mathcal{V}^{*}\mid
T\mathcal{V}\,\rangle \right] -8a\partial_{[\mu}\left(\langle\,
\mathcal{V}\mid
T\mathcal{V}^{*}\,\rangle\xi_{\nu]}\right)\nonumber\\
& \nonumber \\
& +16a\langle\, \mathcal{F}_{\mu\nu}\mid T
\left(\Lambda+\xi^{\rho}A_{\rho}\right)\, \rangle
-\tfrac{c}{8}\xi^{\sigma}\tfrac{1}{\sqrt{\vert
g\vert}}\epsilon_{\sigma\mu\nu\rho}\hat{J}^{\rho}
-c\partial_{[\mu}\langle\, \mathcal{A}_{\nu]}\mid T
\left(\Lambda+\xi^{\rho}\mathcal{A}_{\rho}\right)\rangle\nonumber\\
& \nonumber \\
&
+\tfrac{c}{2}\langle\, \mathcal{F}_{\mu\nu}\mid T\Lambda\,\rangle
+c\langle \mathcal{F}_{\mu\nu}\mid T\xi^{\rho}\mathcal{A}_{\rho}\,\rangle\, ,
\end{align}

\noindent
where $\Lambda$ is the 1-form gauge transformation parameter given in
\eqref{oneformgaugeparameter}. This can be seen to be equal to the desired
result, Eq.~\eqref{commutatorA2}, for $c=-16a$ and $a=-1/2$. We thus obtain
the following supersymmetry variation rule for $B_{\mu\nu}$

\begin{eqnarray}
\delta_{\epsilon}B_{\mu\nu}
& = &
-\tfrac{1}{2}\langle\, \mathfrak{D}_{i}\mathcal{V}\mid
T\mathcal{V}^{*}\, \rangle\,
\bar{\epsilon}_{I}\gamma_{\mu\nu}\lambda^{iI}+\text{c.c.}
\nonumber\\
& & \nonumber \\
& &
-i\langle\, \mathcal{V}\mid T\mathcal{V}^{*}\, \rangle\,
\bar{\epsilon}^{I}\gamma_{[\mu}\psi_{I\nu]}+\text{c.c.}\nonumber\\
& & \nonumber \\
& &
+8\langle\, \mathcal{A}_{[\mu}\mid T\delta_{\epsilon}\mathcal{A}_{\nu]}\, \rangle\,.
\end{eqnarray}

\noindent
The 1-form gauge transformation parameter $\Lambda_{\mu}$ is given
by

\begin{equation}
\Lambda_{\mu}=
2\langle\, \mathcal{V}\mid T\mathcal{V}^{*}\, \rangle\xi_{\mu}
-4\langle\,  \mathcal{A}_{\mu}\mid T(\Lambda+\xi^{\rho}\mathcal{A}_{\rho})
\, \rangle
+\xi^{\rho}B_{\mu\rho}\, .
\end{equation}


\section{World-sheet actions: the vector case}
\label{sec-1braneeffectiveactions}

In this Section we will construct the leading terms of the bosonic part of a
$\kappa$-invariant world-sheet action for the stringy cosmic strings that
couple to the 2-form potentials $B$ that were constructed in Section
\ref{sec-2forms}. Just as in the 0-brane case of Section
\ref{sec-0braneeffectiveactions}, we will construct actions which are
manifestly symplectic invariant.

According to the results of the previous Sections we expect to have strings
which carry charges with respect to each of the ${\rm dim}\, G_{V}$ 2-forms
$B_{A\, \mu\nu}$ that one can define. We define a ${\rm dim}\,
G_{V}$-dimensional charge vector $q^{A}$. Symplectic invariance suggests a
world-sheet action with leading terms

\begin{equation}
\label{stringactionvectorcase}
 S=q^{A}\int d^{2}\sigma\,\langle\mathcal{V}\mid T_A\mathcal{V}^*\rangle
\sqrt{|g_{(2)}|}+cq^{A}\int B_{A}\, ,
\end{equation}

\noindent
where $g_{(2)}$ and $B_{A}$ are the pullbacks of the space-time metric and
2-forms onto the world-sheet, respectively and where $c$ is some normalization
constant that will be fixed later. The tension of the string is given by the
momentum map $\mathcal{P}_A^0$ as given in Eq.~\eqref{momentummap}.

The Wess--Zumino term of this action is, however, not gauge invariant under
the gauge transformation \eqref{gaugetrafoB} and it seems impossible to make
it gauge invariant by adding additional terms to the Wess--Zumino term without
adding more degrees of freedom to the 2-dimensional world-sheet theory.

Actually, the same problem arises in the construction of a $\kappa$-symmetric
world-sheet action for the heterotic superstring in backgrounds with
non-trivial Yang--Mills fields since the NSNS 2-form transforms under
Yang--Mills gauge transformations similar to Eq.~\eqref{gaugetrafoB}. In the
10-dimensional case of strings propagating in backgrounds with non-trivial
Yang-Mills fields the solution to this puzzle lies in the addition of
heterotic fermions to the world-sheet action whose gauge transformations
cancel against the Yang--Mills part of the NSNS 2-form gauge transformation
\cite{Atick:1985iy}. We suggest that a similar effect could be at work here.

If this is the case, then, in checking the invariance under supersymmetry
transformations of the above world-sheet action we must ignore the term
$\langle\, \mathcal{A}_{[\mu}\mid T\delta_{\epsilon}\mathcal{A}_{\nu]}\rangle$
in the 2-form supersymmetry transformation rule. This term should be cancelled
by anomalous terms in the supersymmetry transformations of the world-sheet
spinors. With this proviso we find that the above action preserves half of the
supersymmetries with the projector

\begin{equation}\label{vectorstringprojector}
{\textstyle\frac{1}{2}}(1 +4c\gamma_{01})\epsilon_{I}=0\,
\hspace{.5cm}\text{with}\hspace{.5cm}c=\tfrac{1}{4}\,.
\end{equation}

We will see in the next Section that the stringy cosmic string solutions for
which the above action provides the sources require in order to preserve half
of the supersymmetries exactly the same condition to be satisfied by the
Killing spinor.


\section{Supersymmetric vector strings}
\label{sec-stringsolutions}

Stringy cosmic string solutions of $N=2,d=4$ supergravity coupled to vector
multiplets were found in \cite{Meessen:2006tu}\footnote{Solutions related to
  these by dimensional reduction have been obtained in a 3-dimensional context
  in Ref.~\cite{Howe:1995zm}.}.  They preserve half of the original
supersymmetries and belong to the `null class' of supersymmetric solutions
characterized by the fact that the Killing vector that one can construct from
their Killing spinors is null.  Generically solutions in this class have
Brinkmann-type metrics

\begin{equation}\label{eq:bmmetric}
ds^{2} = 2 du (dv + H du +\hat{\omega})
-2e^{-\mathcal{K}(Z,Z^{*})}dzdz^{*}\, ,
\end{equation}

\noindent
where $\mathcal{K}$ is the K\"ahler potential of the vector scalar manifold
and where $\hat{\omega}$ is determined from the equation

\begin{equation}
(d\hat{\omega})_{\underline{z}\underline{z}^{*}}
=
2i e^{-\mathcal{K}}\mathcal{Q}_{\underline{u}}\, ,
\end{equation}

\noindent
with $\mathcal{Q}_{\mu}$ the pullback of the K\"ahler 1-form connection given
in Eq.~\eqref{eq:K1form}. The complex scalars $Z^{i}$ are functions of $u$ and
$z$.

It is not easy to interpret physically these solutions for a generic
dependence on the null coordinate $u$. When there is no dependence on $u$ we
can take $\hat{\omega}=0$ and the metric is that of a superposition of cosmic
strings (described by $\mathcal{K}$) lying in the direction $u-v$ and
gravitational and electromagnetic waves (described by $H$) propagating along
the same direction.

Setting $H=0$ (which generically requires that we switch off all the
electromagnetic fields) we obtain solutions that only describe cosmic strings.
In order to study the behavior of these solutions under the symmetries of the
theory, it is convenient to express them in an arbitrary system of holomorphic
coordinates, which amounts to the introduction of an arbitrary holomorphic
function $f(z)$ whose absolute value appears in the metric and whose phase
appears in the Killing spinors of the solution

\begin{equation}\label{vectorsolution}
\left\{
\begin{array}{rcl}
ds^{2} & = & 2dudv -2e^{-\mathcal{K}(Z,Z^{*})}|f|^{2} dzdz^{*}\, ,\\
& & \\
Z^{i} & = & Z^{i}(z)\, ,\hspace{1cm} f= f(z)\, ,\\
& & \\
\epsilon_{I} & = & (f/f^{*})^{1/4}\epsilon_{I\, 0}\, ,
\hspace{1cm} \gamma_{z^{*}}\epsilon_{I\, 0}=0\, .
  \end{array}
\right.
\end{equation}

\noindent
If we take $z=x_2+ix_3$ then the condition $\gamma_{z^{*}}\epsilon_{I\, 0}=0$
is equivalent to Eq.~\eqref{vectorstringprojector}.

The holomorphic functions $Z^{i}(z),f(z)$ are assumed to be defined on the
Riemann sphere $\hat{\mathbb{C}}$, but, generically, they will not be
single-valued on it due to the presence of branch cuts. These branch cuts are
to be associated with the presence of cosmic strings just as was done in the
particular case of the $SL(2,\mathbb{R})/U(1)$ special K\"ahler manifold
studied in Refs.~\cite{Bergshoeff:2007aa} and \cite{Bergshoeff:2006jj}.

As a general rule bosonic fields must be single-valued unless they are subject
to a gauge symmetry which forces us to identify as physically equivalent those
configurations which are related by admissible gauge transformations. In the
theories that we are considering the complex scalars $Z^{i}(z)$ do not
transform under any gauge symmetry. Only the global group of isometries
$G_{V}$ of $\mathcal{G}_{ij^{*}}$ acts on them and only a discrete subgroup
$G_{V}(\mathbb{Z})\subseteq Sp(2n_V+2,\mathbb{Z})$ will be a global symmetry
at the quantum level.

In the resulting theories two values of $Z^{i}(z)$ may be considered
equivalent if they are related by a $G_{V}(\mathbb{Z})$ transformation. This
enables one to construct solutions in which the scalars $Z^{i}(z)$ are
multi-valued functions with branch cuts related to the elements of
$G_{V}(\mathbb{Z})$. The source for a branch cut is provided by the
Wess--Zumino term of a cosmic string. This is explained in detail for the
10-dimensional case of the 7-branes in \cite{Bergshoeff:2007aa}.

Next we discuss the emergence of axions related to the presence of Killing
vectors. For every Killing vector $\alpha^{A}k_{A}{}^{i}$ one can always find
an adapted coordinate system $\{Z^{i}\}$ such that the metric
$\mathcal{G}_{ij^{*}}$ does not depend on the real part of the coordinate
$Z^{1}$, say. In this coordinate system
$\alpha^{A}k_{A}{}^{i}\partial_{i}=\partial_{1}$ and the isometries generated
by it act as constant shifts of $Z^{1}$ by a real constant:

\begin{equation}
\label{axions}
\delta Z^{1}=c\in\mathbb{R}\, .
\end{equation}

\noindent This transformation only acts on the real part of
$Z^{1}$, $\chi^{1}$, which is, then, what it is sometimes meant by
an axion: a real scalar field with no non-derivative couplings to
the other scalars and with a shift symmetry\footnote{A more
precise definition would require $\chi^{1}$ to be a pseudoscalar
too. Actually, the real and imaginary parts of the complex scalars
in $N=2,d=4$ vector supermultiplets have different parities, but,
in a general model with arbitrary coordinates one should look at
the couplings to the vector fields to determine the parity of
$\chi^{1}$.

On the other hand, the action of $N=2,d=4$ supergravity indicates
that the axions must appear in $\Re{\rm e}\,
\mathcal{N}_{\Lambda\Sigma}$, which couples to the parity-odd term
$F^{\Lambda} \wedge F^{\Sigma}$. Under symplectic transformations
$\left(
    \begin{array}{cc}
     1 & B\\
     0 & 1\\
    \end{array}
  \right)$ $\Re{\rm e}\, \mathcal{N}$ is shifted to $\Re{\rm e}\,
\mathcal{N}+B$, as one expects from axions. This suggests another
possible characterization of axions: $\chi^{1}$ is an axion if its
shifts are embedded in the Abelian subgroup of symplectic
transformations of the form $\left(
    \begin{array}{cc}
     1 & B\\
     0 & 1\\
    \end{array}
  \right)$. }

It is clear that we can, in principle, define as many different
axion fields as there are independent Killing vectors
\footnote{However, they cannot be used simultaneously, since we
can only use simultaneously adapted coordinates for commuting
isometries.}, i.e.~$\mathrm{dim}\, G_{V}$, i.e.~as many as
2-forms, which can be understood as their duals. Their (both those
of the axions and 2-forms) equations of motion are not necessarily
independent, though, and they will satisfy a number of
constraints, as discussed before, and, at most, there can be
$2n_{V}$ independent axions.

We now discuss the properties of the cosmic string solutions in a
local neighborhood of the location $z_0$ in the transverse space
of a cosmic string. Infinitesimally the transformation of the
scalars $Z^i$ when going around $z_0$ is given by
Eq.~\eqref{deltaZ}. In some coordinate basis, the transformation
will only be an axion shift.

Besides the scalars $Z^i$ also the Killing spinors $\epsilon_I$ will undergo
transformations when going around the cosmic string at $z_0$. This is because
when the scalars transform as in Eq.~\eqref{deltaZ} the K\"{a}hler potential
transforms as

\begin{equation}
\mathcal{K}(Z^{\prime},Z^{\prime *}) = \mathcal{K}(Z,Z^{*})
+\lambda_{\alpha}(Z) +\lambda^{*}_{\alpha}(Z^{*})\, .
\end{equation}

\noindent
From the fact that the Killing spinor $\epsilon_I$ has K\"{a}hler weight 1/2
it then follows that

\begin{equation}
\epsilon_{I}(z) \rightarrow
e^{\frac{1}{4}[\lambda_{\alpha}-\lambda^{*}_{\alpha}]
+\frac{i}{2}\varphi_{\alpha}}\epsilon_{I}(z)\, ,
\end{equation}

\noindent when going around $z_0$. The phases $\varphi_{\alpha}$
relate to the fact that in general the spinors transform under the
double cover of $G_{V}$\footnote{One can even include yet another
phase factor in the transformation rule for the Killing spinors
which incorporates the fact that $\epsilon_I$ may come back to
itself up to a sign, i.e.~one can include nontrivial spin
structures.}. The Killing spinor $\epsilon_I$ is defined in terms
of the holomorphic function $f(z)$ via
Eqs.~\eqref{vectorsolution}. The monodromy of $f$ when going
around $z_{0}$ must be

\begin{equation}
f(z)\rightarrow
e^{\lambda_{\alpha}[Z(z)]+i\varphi_{\alpha}}f(z)\,.
\end{equation}

The cosmic string solutions contain information about the moduli space of the
theory, i.e.~the space of inequivalent values for $Z^i$. The classical moduli
space is defined by the requirement

\begin{equation}
\label{classicalmodulispace}
    \rm Im\,\mathcal{N}_{\Lambda\Sigma}<0\,,
\end{equation}

\noindent
in order that the kinetic terms of the 1-forms have the right sign in the
action \eqref{action}. The zeros of the polynomial $\delta Z^i=\alpha^A
k_{A}{}^{i}$ which belong to the space \eqref{classicalmodulispace} (or
possibly on the boundary thereof) are fixed points of the monodromy
 and therefore comprise the loci of the cosmic strings in the
quantum moduli space:

\begin{equation}
\label{quantummodulispace}
    \{Z^i\,\vert\,\rm Im\,\mathcal{N}_{\Lambda\Sigma}<0\}/G_V(\mathbb{Z})\,.
\end{equation}

Drawing from the analogy with the $SL(2,\mathbb{R})/U(1)$ case studied in
\cite{Bergshoeff:2006jj} one can expect all physical properties of globally
well-defined stringy cosmic string solutions to be mapped into geometrical
properties of the space \eqref{quantummodulispace}. Such properties are the
total mass, possible deficit angles at the sites of the cosmic strings, orders
of monodromy transformations (the number of times the same monodromy has to be
applied in order to equal the identity), etc.  Here we will not attempt to
work out the global properties of these solutions, since they are strongly
model-dependent.

In the $SL(2,\mathbb{R})/U(1)$ case one could have derived all geometrical
properties of the quantum moduli space $SL(2,\mathbb{Z})\backslash
SL(2,\mathbb{R})/U(1)$  by studying the globally well-defined
supersymmetric stringy cosmic string solutions. It is therefore natural to ask
the question whether this is generally true, i.e.~whether (some class of)
quantum moduli spaces of Calabi--Yau reduced supergravities can be obtained by
studying the properties of the stringy cosmic string solutions.

We leave this for a future investigation.


\section{The 2-forms: the hyper case}
\label{sec-hyper2forms}

If we consider $N=2$, $d=4$ supergravity with general matter
couplings, we can have apart from the complex scalars in the
vector multiplets $4n_H$ real scalars when coupling gravity to
$n_H$ hypermultiplets. In the following we repeat the program of
introducing 2-forms in order to dualize the hyperscalars which
parameterize the Noether currents of some isometry group of the
quaternionic K\"{a}hler manifold. We first construct the Noether
currents, dualize them and subsequently construct the
supersymmetry transformation rule for the dual 2-forms. For the subset of commuting isometries a similar program has been performed in \cite{Theis:2003jj} where also actions for the dualized scalars are given.

\subsection{The Noether current}

The transformations we are dealing with are just the isometries of
the quaternionic K\"{a}hler manifold that we write in the form

\begin{equation}
\delta q^{u}=\alpha^{A} k_{A}{}^{u}(q)\, ,
\end{equation}

\noindent where $k_{A}{}^{u}$ are the components of the Killing
vectors $k_A=k_A{}^u\partial_u$ that generate the isometry group
$G_{H}$ of $\mathsf{H}_{uv}$. The parameters $\alpha^{A}$ are real
parameters.

Associated to each of the isometries we can define a momentum
map\footnote{Momentum maps play a crucial role in the gauging of
the isometries. It is therefore interesting to note that the
mathematics which governs the 2-forms is similar to that used in
gauged matter coupled $N=2$, $d=4$ supergravity.}
$\mathsf{P}_{AI}{}^{J}$ defined by the equation

\begin{equation}\label{defmomentummaps}
\mathfrak{D}_{u}\mathsf{P}_{AI}{}^{J}=
-\mathsf{J}_{I}{}^{J}{}_{uv}k_{A}{}^{v}\, ,
\end{equation}

\noindent where $\mathsf{J}_{I}{}^{J}{}_{uv}$ is the triplet
complex structures of the quaternionic-K\"aher manifold.

Following \cite{Aoyama:2005hb} we write the triplet of complex
structures $\mathsf{J}_{I}{}^{J}{}_{uv}$ in terms of the Quadbeins
as follows

\begin{equation}\label{complexstructure}
    \mathsf{J}_{I}{}^{J}{}_{uv}=\tfrac{i}{2}(\sigma_x)_I{}^{J}\,\mathsf{J}^x{}_{uv}\hspace{.5cm}\text{with}\hspace{.5cm}
    \mathsf{J}^x\,{}^u{}_v=-i\mathsf{U}^{\alpha
    I}{}_v(\sigma_x)_I{}^J\mathsf{U}_{\alpha J}{}^u\,,
\end{equation}

\noindent where the $\sigma_x$, $x=1,2,3$, are the three Pauli
matrices. We will often write $\mathsf{P}_{I}{}^{J}\equiv
\alpha^{A}\mathsf{P}_{AI}{}^{J}$.

The Noether current associated to the these isometries, which do
not act on the vector fields, is just

\begin{equation}
\label{eq:hyperNC} J_{N}^{\mu} =\delta q^{u} \frac{1}{\sqrt{|g|}}
\frac{\partial\mathcal{L}}{\partial(\partial_{\mu} q^{u})}
=4\mathsf{H}_{uv}\partial^{\mu} q^{v}\delta q^{u}\, ,
\end{equation}

\noindent and satisfies $\nabla_{\mu}J_N^{\mu}=0$.

\subsection{Dualizing the Noether current}

Since the isometries of the quaternionic K\"{a}hler manifold do
not act on the vectors of the theory they are symmetries of the
action and there will be no anomalous contribution to the Noether
current such as $\hat J$ which we encountered when discussing the
isometries of the special K\"{a}hler manifold. We can thus
immediately define the gauge-invariant 3-form field strength $H$
via

\begin{equation}
H=dB=\star J_{N}\, ,
\end{equation}

\noindent where $H=\alpha^A H_A$ and $B=\alpha^A B_A$.

\subsection{The 2-form supersymmetry transformation}

We know that, since $B$ is defined by $dB=\star J_{N}$, the
commutator of two supersymmetry variations on $B$ must close into
the algebra \eqref{commutatoralgebra}, i.e.~it must lead to the
commutator

\begin{equation}
\label{commutatorA2} [\delta_{\eta},\delta_{\epsilon}]B_{\mu\nu}
=\xi^{\rho}\tfrac{1}{\sqrt{\vert
g\vert}}\epsilon_{\rho\mu\nu\sigma}J_{N}{}^{\sigma}
+2\partial_{[\mu}\left(\Lambda_{\nu]}-\xi^{\rho}B_{\nu]\rho}\right)\,.
\end{equation}

\noindent In order to achieve this, we make the following Ansatz
for the supersymmetry variation of the 2-form (up to second order
in fermions)

\begin{eqnarray}
\delta_{\epsilon} B_{\mu\nu} & = & a
\mathsf{P}_{I}{}^{J}\bar\epsilon^{I}\gamma_{[\mu}\psi_{J\vert\nu]}
+\mathrm{c.c.} \nonumber \\
\nonumber\\
& & +b \mathsf{U}_{\alpha J}{}^{u} \mathfrak{D}_{u}
\mathsf{P}_I{}^{J}\bar\epsilon^{I}\gamma_{\mu\nu}\zeta^\alpha
+\mathrm{c.c.}\, ,
\end{eqnarray}

\noindent where $a$ and $b$ are arbitrary complex constants.

Evaluating the commutator and assuming that $a$ and $ib$ are real
parameters we obtain

\begin{eqnarray}
[\delta_{\eta},\delta_{\epsilon}]B_{\mu\nu} & = & -\tfrac{3}{2}i b(\star
dq^{w})_{\mu\nu\rho} \xi^{\rho}\mathsf{H}_{vw}\delta q^{v}
\nonumber \\
\nonumber \\
& & +\tfrac{3}{2}i b\mathsf{J}{}_I{}^{K}{}_{vw}\delta q^{v}
\partial_{[\nu}q^{w} X_{\mu]}{}_{K}{}^{I}
\nonumber \\
\nonumber \\
& &
+2\partial_{[\mu}\left(\Lambda_{\nu]}-\xi^{\rho}B_{\nu]\rho}\right)
-a\mathsf{J}_{I}{}^{K}{}_{vw}\delta
q^{v}\partial_{[\nu}q^{w}X_{\mu]K}{}^{I}\, ,
\end{eqnarray}

\noindent where we have defined the matrix of vector fields

\begin{equation}
X_{\mu I}{}^{J}\equiv -\bar{\eta}^{J}\gamma_{\mu}\epsilon_{I}
-\bar{\eta}_{I}\gamma_{\mu}\epsilon^{J}\, ,
\end{equation}

\noindent and where the gauge parameter $\Lambda_{\mu}$ is given by
\begin{equation}
\Lambda_{\mu}=
-\tfrac{a}{2}X_{J}{}^{I}{}_{\mu}\mathsf{P}_{I}{}^{J}+\xi^\rho
B_{\mu\rho}.
\end{equation}

\noindent Next we choose $a=\tfrac{3}{2}ib$ and we are left with

\begin{equation}
[\delta_{\eta},\delta_{\epsilon}]B_{\mu\nu}
 = -\tfrac{3}{2}ib(\star dq^{w})_{\mu\nu\rho}
\xi^{\rho}\mathsf{H}_{vw}\delta q^{v} +
2\partial_{[\mu}\left(\Lambda_{\nu]}-\xi^{\rho}B_{\nu]\rho}\right)\,
.\\\nonumber
\end{equation}

\noindent If we compare this expression with
Eq.~\eqref{commutatorA2} using Eq.~\eqref{eq:hyperNC} we read off
that $ib=-\tfrac{8}{3}$, so that $a=-4$.

The supersymmetry transformation of the 2-forms dual to the
hyperscalars parameterizing the Noether current \eqref{eq:hyperNC}
is thus

\begin{eqnarray}
\delta_{\epsilon} B_{\mu\nu} & = & -4
\mathsf{P}_I{}^{J}\bar{\epsilon}^{I}\gamma_{[\mu}\psi_{J\vert\nu]}
+\mathrm{c.c.}\nonumber \\
\nonumber\\
& & +{\textstyle\frac{8i}{3}} \mathsf{U}_{\alpha J}{}^{u}
\mathfrak{D}_{u}
\mathsf{P}_{I}{}^{J}\bar{\epsilon}^{I}\gamma_{\mu\nu}\zeta^{\alpha}
+\mathrm{c.c.}\, ,\label{susytrafohyper2form}
\end{eqnarray}

\noindent and the 2-form gauge parameter $\Lambda_{\mu}$ is given
by

\begin{equation}
\Lambda_{\mu}= 2X_{J}{}^{I}{}_{\mu}\mathsf{P}_{I}{}^{J}+\xi^\rho
B_{\mu\rho}.
\end{equation}


\section{World-sheet actions: the hyper case}

Stringy cosmic strings in the hyper case are strings electrically
charged under the 2-forms $B$ constructed in Section
\ref{sec-hyper2forms}. In this Section we will construct the
bosonic part of the string effective action, which preserves half
of the supersymmetries of the theory. In analogy with the Ansatz
that we made for the strings in the vector case we again express
the tension of the string in terms of the momentum maps. We make
the following Ansatz

\begin{equation}\label{eq:actionhyper}
S=\int d^2\sigma \mathcal T_1\,\sqrt{|g_{(2)}|}+c\,q^A
\int B_{A},
\end{equation}
where $c$ is some real number which will be fixed later. The
tension is given by

\begin{equation}
 \mathcal T_1=\sqrt{(\mathsf{P}^x)^2}\hspace{.5cm}\text{where}\hspace{.5cm}\mathsf{P}^x=\alpha^A \mathsf{P}^x{}_A
 \hspace{.5cm}\text{with}\hspace{.5cm}\mathsf{P}_I{}^J=\tfrac{i}{2}\mathsf{P}^x(\sigma_x)_I{}^J
\end{equation}

\noindent and in taking the square we sum over $x=1,2,3$.

Performing a supersymmetry variation of the action
\eqref{eq:actionhyper} using the transformation rules
\eqref{eq:susytranse}, \eqref{eq:susytransq} and
\eqref{susytrafohyper2form} we find that the string action
preserves half of the supersymmetries with a projector given by

\begin{equation}\label{eq:projection}
\Pi_I{}^J=\tfrac{1}{2}(\delta_I{}^J-{\textstyle\frac{8ci}{\sqrt{(\mathsf{P}^x)^2}}}\mathsf{P}_I{}^{J}\gamma_{01}),\qquad\Pi_I{}^J\epsilon^I=0,
\hspace{.5cm}\text{where $c=-\tfrac{1}{4}$}.
\end{equation}

An important distinction with the analogous string action
constructed in Section \ref{sec-1braneeffectiveactions} is that in
the present case the Wess--Zumino term is gauge invariant up to a
total derivative whereas in the case of strings coupled to 2-forms
dual to vector scalars the Wess--Zumino term is not by itself
gauge invariant, cf. the discussion below
Eq.~\eqref{stringactionvectorcase}. In fact one may consider the
action \eqref{eq:actionhyper} as the first example of a 1/2 BPS
$(d-3)$-brane action which is well-defined (at the bosonic level)
for all possible $(d-2)$-form potentials. In the
$d=10$-dimensional situation only the brane actions related to
the D7-branes are well understood. For the other 8-forms which
couple to the Q7-branes of \cite{Bergshoeff:2007aa} there are
still open problems regarding a proper understanding of the
world-volume dynamics. The fact that in the particular case of the
hyperstrings we can construct well-defined actions supports the
idea that in general one can treat all isometries of any scalar
sigma model in any supergravity on an equal footing (provided they
pertain to be discrete isometries of the quantum moduli space).
This suggests that in order to find the full spectrum of 1/2 BPS
states one best considers the same supergravity theory in various
coordinate systems in which these isometries take on a simple
form.

\section{Supersymmetric hyperstrings}\label{hypersolutions}

In Ref.~\cite{Huebscher:2006mr} it was shown that the c-map
transforms supersymmetric stringy cosmic string solutions of the
vector scalar manifold into supersymmetric stringy cosmic string
solutions of the hyperscalar manifold. The latter belong to the
timelike class of supersymmetric solutions characterized by the
fact that the Killing vector that one can construct from the
Killing spinors of the solution is timelike.  The metric for this
class of solutions (for vanishing vector multiplets) takes the
following form

\begin{equation}
ds^{2}= dt^{2} -\gamma_{\underline{m}\underline{n}}dx^{m}dx^{n}\, .
\end{equation}

The 3-dimensional spatial metric
$\gamma_{\underline{m}\underline{n}}$ (or its Dreibeins
$V^{x}{}_{\underline{m}}$) is related to the hyperscalars
$q^{u}(x)$ by two conditions. The first condition is

\begin{equation}
\label{eq:hyperkse4}
V_{x}{}^{\underline{m}}\,\partial_{\underline{m}}q^{u}\
\mathsf{U}^{\alpha J}{}_{u}\,(\sigma_{x})_{J}{}^{I}\; =\; 0\, ,
\end{equation}

\noindent and the second condition reads, in a given $SU(2)$ and
Lorentz gauge,

\begin{equation}
\label{eq:connectionembedding}
\varpi_{\underline{m}}{}^{xy} = \varepsilon^{xyz}
\mathsf{A}^{z}{}_{u}\ \partial_{\underline{m}}q^{u}\, ,
\end{equation}

\noindent where $\varpi_{\underline{m}}{}^{xy}$ is the spin
connection 1-form of the 3-dimensional metric and
$\mathsf{A}^{z}{}_{u} \partial_{\underline{m}}q^{u}$ is the
pullback of the $SU(2)$ connection of the quaternionic-K\"ahler
manifold parameterized by the scalars $q^{u}$. In the gauge in
which Eq.~(\ref{eq:connectionembedding}) holds the Killing spinors
take the form

\begin{align}
& \epsilon_{I}=\epsilon_{I\, 0}, \hspace{.5cm}
\Pi^{x}{}_{I}{}^{J}\ \epsilon_{J\, 0} =0\hspace{.5cm}
\text{with}\hspace{.5cm}\Pi^{x}{}_{I}{}^{J} \;\equiv\;
{\textstyle\frac{1}{2}} [\ \delta_{I}{}^{J}\ -\ \gamma^{0(x)}\
(\sigma_{(x)})_{I}{}^{J}\ ]\label{constraint2}
\end{align}

\noindent where the notation $(x)$ in \eqref{constraint2} means
that $x$ is not summed over so the constraints are imposed for
each non-vanishing component of the $SU(2)$ connection.

We now repeat for the hyperscalars parameterizing a quaternionic
K\"{a}hler manifold with isometry group $G_{H}$ the discussion of
Section \ref{sec-stringsolutions}. The fields will only depend on
two spatial coordinates ($x^{1}$ and $x^{2}$, say, that can always
be combined into a complex coordinate $z$) which parameterize the
transverse space of the cosmic string. The metric will take the
form

\begin{equation}
ds^{2}= dt^{2}-(dx^{3})^{2}
-2e^{\Phi(z,z^{*})}dzdz^{*}\, ,
\end{equation}

\noindent
and the hyperscalars will be real functions $q^{u}(z,z^{*})$. A convenient
Dreibein basis is

\begin{equation}
\hat{V}^{3}=dx^{3}\, , \hspace{.5cm} \hat{V}^{z} = V dz\, ,
\hspace{.5cm} \hat{V}^{z^{*}} = V^{*} dz^{*}\, , \hspace{.5cm}
|V|^{2} = e^{\Phi(z,z^{*})} \, .
\end{equation}

\noindent In this Dreibein basis the supersymmetry conditions
Eqs.~(\ref{eq:hyperkse4}) and (\ref{eq:connectionembedding}) take
the respective form

\begin{eqnarray}
\label{eq:embeddingagain1}
 \mathsf{U}^{\alpha 2}{}_{u}
\partial_{\underline{z}}q^{u}
=
\mathsf{U}^{\alpha 1}{}_{u}
\partial_{\underline{z}^*}q^{u}
&  = &  0\, , \\
& & \nonumber \\
\label{eq:embeddingagain2}
\varpi_{\underline{z}}{}^{zz^{*}} & = &
\mathsf{A}^{3}{}_{u}\ \partial_{\underline{z}}q^{u}\, ,\\
& & \nonumber \\
\label{eq:embeddingagain3}
\mathsf{A}^{1}{}_{u}\ \partial_{\underline{m}}q^{u} =
\mathsf{A}^{2}{}_{u}\ \partial_{\underline{m}}q^{u} & = & 0\, .
\end{eqnarray}

The Killing spinors of these solutions, in this basis, are given
by

\begin{equation}\label{constraint3}
\epsilon_{I}=\epsilon_{I\, 0}\, , \hspace{.5cm}
\Pi^{3}{}_{I}{}^{J}\ \epsilon_{J\, 0} =0\, .
\end{equation}

It can be shown that in this gauge the pullbacks
of the complex structures $\mathsf{J}^1$ and $\mathsf{J}^2$ vanish
while $\mathsf{J}^3$ remains nonzero and one recovers the projection
operator Eq.~(\ref{eq:projection}).
As in the case of the vector scalars, it is convenient to work in
a more general coordinate system in which the metric takes the
form

\begin{equation}
ds^{2}= dt^{2}-(dx^{3})^{2}
-2e^{\Phi(z,z^{*})}|f|^{2}dzdz^{*}\, ,
\end{equation}

\noindent where $f(z)$ is a holomorphic function. The
supersymmetry conditions, Eqs.~(\ref{eq:embeddingagain1}) and
(\ref{eq:embeddingagain3}), do not change and
Eq.~(\ref{eq:embeddingagain2}) is still satisfied \textit{with the
old spin connection}. If the new spin connection is computed with
respect to the new frame

\begin{equation}
\hat{V}^{3}=dx^{3}\, ,
\hspace{.5cm}
\hat{V}^{z} = Vf^{*} dz\, ,
\hspace{.5cm}
\hat{V}^{z^{*}} = V^{*}f dz^{*}\, ,
\end{equation}

\noindent
then, we find that

\begin{equation}
\varpi_{\underline{z}}{}^{zz^{*}}=
\varpi_{\underline{z}}{}^{zz^{*}}{}_{\rm old}+
\partial_{\underline{z}}\log{f}\, ,
\end{equation}

\noindent
and then the Killing spinors take the form

\begin{equation}
\epsilon_{I}=e^{\frac{1}{2} \log (f/f^{*})\gamma^{03}}\epsilon_{I\, 0}\, ,
\end{equation}

\noindent the constant spinor $\epsilon_{I\, 0}$ obeying the same
constraints as above, Eqs.~\eqref{constraint3}. These same
constraints allow us to rewrite it in the equivalent form

\begin{equation}
\epsilon_{I}=\exp{\{{\textstyle\frac{1}{2}} \log
(f/f^{*})\sigma_{3}\}}_{I}{}^{J}\epsilon_{J\, 0}\, .
\end{equation}

The multi-valuedness of the Killing spinors $\epsilon_I$ of these
solutions is related to the $U(1) \subset SU(2)$ gauge
transformation where the $U(1)$ subgroup is associated to the
non-vanishing component
$\mathsf{A}^{3}{}_{u}\partial_{\underline{z}}q^u$ of the $SU(2)$
connection pulled back on the space-time. The transformations of
the Killing spinors determine the monodromy properties of the
holomorphic function $f$ similarly to what happens in the case of
the vector scalars.


\section{Conclusions}
\label{sec-conclusions}

In this paper we have shown how, consistent with the supersymmetry
algebra, the standard set of bosonic fields of $N=2,d=4$
supergravity coupled to vector and hypermultiplets can be extended
to include $n_{V}+1$ additional ``magnetic''  vector fields and
$\mathrm{dim}\, G_{V}$ 2-form fields dual to vector multiplet
scalars, as well as $\mathrm{dim}\, G_{H}$ 2-form fields dual to
hypermultiplet scalars. These fields couple, respectively, to
magnetic 0-branes (black holes) and cosmic strings for which there
are  well-known classical solutions that we have reviewed. They
are necessary to construct $\kappa$-symmetric effective
world-volume actions for these solutions. We have studied the
construction of these actions in a symplectic-covariant form and
checked that their supersymmetry to lowest order precisely leads
to the 1/2 BPS condition one expects for these solutions.

One possible extension is based on the idea that there may also be
3- and 4-form potentials (also known as deformation potentials and
top-form potentials, respectively) unrelated by duality to any of
the standard fields of the theory and which do not carry any
(continuous) degree of freedom. Deformation and top-form
potentials have been found and studied in 10-dimensional
supergravities
\cite{Bergshoeff:1998re,Riccioni:2004nr,Bergshoeff:2005ac,Bergshoeff:2006qw}.
These potentials can be associated with higher-dimensional objects
such as  domain-walls and space-time-filling branes. For a recent
derivation of the representations of these potentials for maximal
supergravity from a Kac-Moody point of view, see
\cite{Riccioni:2007au,Bergshoeff:2007qi}. It would be very
interesting to carry out a similar analysis in the $N=2,d=4$
theories. For the cases that the special K\"ahler manifold
corresponds to a coset geometry the representations of these
potentials again follow from a Kac-Moody approach
\cite{Bergshoeff:2007vb}. Alternatively, some of the deformation and
top-form potentials should be related by dimensional reduction to
those of minimal $d=5$ supergravity, which have recently been
constructed in \cite{Gomis:2007gb}\,\footnote{All the deformation
and top-form potentials of minimal $d=5$ supergravity will give
rise to top-form potentials in 4 dimensions. However, in general,
not all these potentials can be obtained from a higher-dimensional
theory, the best-known example being the RR 9-form potential of
$N=2A,d=10$ supergravity.}. Further, the deformation potentials
carry a great deal of information about possible gaugings or
massive deformations (hence the name) of the supergravity theory.
It would be interesting to work these things out in detail for the
$N=2,d=4$ theories.

There is yet another interesting connection between gauged
supergravity and the $(d-2)$-form potentials that we have studied
here which is worth exploring. It is known that if one performs
generalized (Scherk-Schwarz) dimensional reductions associated to
one isometry of a sigma model metric in $d$ space-time dimensions,
one gets gauged supergravities
\cite{Bergshoeff:1997mg,Lavrinenko:1997qa,Kaloper:1998kr,Hull:1998vy,Cowdall:2000sq,Hull:2002wg,Bergshoeff:2002mb}
in $d-1$ space-time dimensions. Locally, these generalized
dimensional reductions can be interpreted as reductions in the
background of the $(d-3)$ brane that would couple to the
$(d-2)$-form potential dual to the Noether current associated to
the isometry used in the reduction
\cite{Meessen:1998qm,Bergshoeff:2002mb,Bergshoeff:1996ui}. After
reduction, in the transverse direction, the $(d-3)$ branes become
domain-wall solutions in the reduced theory and should couple to
deformation potentials directly obtainable from the $(d-2)$-form
potentials of the original theory.

In particular, in the case at hand, we should be able to perform
explicit generalized dimensional reductions using isometries of
the special K\"ahler manifold in a way consistent with all the
symmetries of the theory (as it was done in \cite{Meessen:1998qm})
down to 3 dimensions, obtaining gauged 3-dimensional
supergravities on the one hand. On the other hand, we should be
able to relate the deformation parameters that appear in 3
dimensions with deformation potentials (i.e.~2-form potentials)
which can be obtained from the 4-dimensional 2-form potentials
that we have obtained here. At the same time one should be able to
relate the 4-dimensional cosmic string solutions to the
3-dimensional domain-wall solutions. Similar relations between the
5- and 4-dimensional theories must exist. Work on these subjects
is in progress.


\section*{Acknowledgments}

We would like to thank Patrick Meessen for many useful
conversations.  This work has been supported in part by the Spanish
Ministry of Science and Education grants FPA2006-00783 and
PR2007-0073 (TO), the Comunidad de Madrid grant HEPHACOS P-ESP-00346
and by the EU Research Training Network \textit{Constituents,
  Fundamental Forces and Symmetries of the Universe} MRTN-CT-2004-005104 in
which E.B.~and J.H.~are associated to the University of Utrecht and M.H.~and T.O.~are
associated to the University Aut\'onoma de Madrid. J.H.~was supported by a
Breedte Strategie grant of the University of Groningen and wishes to thank the
Instituto de F\'{i}sica Te\'{o}rica of the Universidad Aut\'{o}noma de Madrid
for its hospitality and financial support. M.H.~is supported by the Spanish
Ministry of Science and
Education grant FPU AP2004-2574. T.O.~would like to thank the center
for Theoretical Physics of the University of Groningen and the Stanford
Institute for Theoretical Physics for its hospitality in, respectively, the
earliest and last stages of this work and M.M.~Fern\'andez for her continuous
support.

\appendix


\section{Conventions}
\label{gammaspinor}

The signature is mostly minus. Flat tangent space indices are denoted by lower
case Latin indices $a$ whose values are $a=0,1,2,3$. Curved space-time indices
are denoted by lower case Greek indices $\mu$ whose values are
$\mu=t,\underline{1},\underline{2},\underline{3}$. The tangent space
Levi-Civit\`{a} symbol is taken to be $\epsilon^{0123}=-\epsilon_{0123}=1$.
The curved Levi-Civit\`{a} tensor whose indices are lowered with the metric is
taken to be

\begin{equation}
    \epsilon^{\mu_1\cdots\mu_4}=\sqrt{\vert g\vert}e^{\mu_1}_{a_1}\cdots e^{\mu_4}_{a_4}\epsilon^{a_1\cdots
    a_4}\,,
\end{equation}

\noindent
where $e^{\mu}_a$ is the inverse Vielbein. The Hodge
dual of a $k$-form $\omega$ is defined to be

\begin{equation}
    (*\omega)_{\mu_1\cdots\mu_{d-k}}=\frac{1}{k!\,\sqrt{\vert g\vert}}
    \epsilon_{\mu_1\cdots\mu_{d-k}\nu_{1}\cdots\nu_k}\omega^{\nu_{1}\cdots\nu_k}\,.
\end{equation}

\noindent
The Riemann tensor is defined by
$R_{\mu\nu\rho}^{\quad\;\;\sigma}=\partial_{\mu}\Gamma_{\nu\rho}^{\sigma}+\cdots$.

We work in the Majorana representation which in signature $(+---)$ has all the
gamma matrices purely imaginary,

\begin{equation}
\gamma_a^*=-\gamma_a\, .
\end{equation}

\noindent
The anticommutator is

\begin{equation}
\{\gamma_{a},\gamma_{b}\}= +2\eta_{ab}\, .
\end{equation}

\noindent
The chirality matrix is defined by

\begin{equation}
\gamma_{5}\equiv -i\gamma^{0}\gamma^{1}\gamma^{2}\gamma^{3}
={\textstyle\frac{i}{4!}} \epsilon_{abcd}
\gamma^{a}\gamma^{b}\gamma^{c}\gamma^{d}\, .
\end{equation}

\noindent
With this chirality matrix, we have the identity

\begin{equation}
\label{eq:dualgammaidentityind4}
\gamma^{a_{1}\cdots a_{n}}
=\frac{(-1)^{\left[n/2\right]}i}{(4-n)!} \epsilon^{a_{1}\cdots
a_{n}b_{1}\cdots b_{4-n}} \gamma_{b_{1}\cdots b_{4-n}}
\gamma_{5}\, ,
\end{equation}

\noindent
where $\left[n/2\right]$ is the highest integer less than or equal to $n/2$.
The following two gamma matrix identities are used in the text

\begin{align}
    \gamma_{\mu\nu}\gamma_{\rho}&=\gamma_{\mu\nu\rho}+\gamma_{\mu}g_{\nu\rho}-\gamma_{\nu}g_{\mu\rho}\,,\\
    \gamma_{\mu\nu}\gamma_{\rho\sigma}&=i\epsilon_{\mu\nu\rho\sigma}\gamma_5-2g_{\mu[\rho}g_{\sigma]\nu}-2\gamma_{\mu[\rho}g_{\sigma]\nu}+2g_{\mu[\rho}\gamma_{\sigma]\nu}\,.\label{gammaidentity2}
\end{align}

We use 4-component chiral spinors $\chi$ whose chirality is related to the
position of the $SU(2)$ index $I$ or the position of the $Sp(2n_H)$ index
$\alpha$,

\begin{eqnarray}
\gamma_{5}\chi_{I}=-\chi_{I}\,,\quad & \gamma_{5}\chi^{I}=\chi^{I}\,, \\
\gamma_{5}\chi_{\alpha}=-\chi_{\alpha}\,,\quad &
\gamma_{5}\chi^{\alpha}=\chi^{\alpha}\,.
\end{eqnarray}

\noindent
The position of the $SU(2)$ index I and of the $Sp(2n_H)$ index $\alpha$ is
raised and lowered under complex conjugation

\begin{equation}
    \chi_{I}^{*}=\chi^{I}\hspace{.5cm}\text{and}\hspace{.5cm}\chi_{\alpha}^{*}=\chi^{\alpha}\,.
\end{equation}

\noindent
The conjugated spinor is taken to be

\begin{equation}
{\bar\chi}_I=i(\chi^I)^{\dagger}\gamma_{0}
\hspace{.5cm}\text{and}\hspace{.5cm}
{\bar\chi}_{\alpha}=i(\chi^{\alpha})^{\dagger}\gamma_{0}\,.
\end{equation}

\noindent
The spinors are anticommuting and we take the convention that they do not
change their order under complex conjugation. We have the following property
for spinor bilinears

\begin{equation}
\bar\chi_1\gamma^{\mu_1\cdots\mu_n}\chi_2=(-1)^{\left[(n+1)/2\right]}\bar\chi_2\gamma^{\mu_1\cdots\mu_n}\chi_1\,,
\end{equation}

\noindent
where $\chi_1$ and $\chi_2$ are arbitrary spinors.


\section{Special K\"ahler Geometry}
\label{sec-specialgeometry}

A K\"ahler manifold $\mathcal{M}$ is a complex manifold with coordinates $Z^i$
and $(Z^i)^*=Z^{*\,i^*}$ whose K\"{a}hler 2-form $\mathcal{J}$ is closed. The
K\"{a}hler 2-form is then locally given by $\mathcal{J}=d\mathcal{Q}$ with
$\mathcal{Q}$ the K\"{a}hler connection 1-form. Both the metric and the
K\"{a}hler connection 1-form can be expressed in terms of the K\"{a}hler
potential $\mathcal{K}$ as follows

\begin{align}
& ds^{2} = 2 \mathcal{G}_{ii^{*}}\ dZ^{i}dZ^{*\, i^{*}}
\hspace{.5cm}\text{with}\hspace{.5cm}\mathcal{G}_{ii^{*}} =
\partial_{i}\partial_{i^{*}}\mathcal{K}\, , \\
& \nonumber\\
& \mathcal{Q} \equiv (2i)^{-1}(dZ^{i}\partial_{i}\mathcal{K} -
dZ^{*\, i^{*}}\partial_{i^{*}}\mathcal{K})\, . \label{eq:K1form}
\end{align}

\noindent
The non-vanishing components of the Levi-Civit\`a
connection on a K\"ahler manifold are given by

\begin{equation}
\label{eq:KCCChrisSymb}
\Gamma_{jk}{}^{i} =
\mathcal{G}^{ii^{*}}\partial_{j}\mathcal{G}_{i^{*}k}\, ,
\hspace{1cm}
\Gamma_{j^{*}k^{*}}{}^{i^{*}}  =
\mathcal{G}^{i^{*}i}\partial_{j^{*}}\mathcal{G}_{k^{*}i} \, .
\end{equation}

The K\"ahler potential is not unique. It is defined up to K\"ahler
transformations,

\begin{equation}
\label{eq:Kpotentialtransfo}
\mathcal{K}(Z,Z^{*})\rightarrow\mathcal{K}(Z,Z^{*})
+\lambda(Z)+\lambda^{*}(Z^*)\,,
\end{equation}

\noindent
where $\lambda$ is any holomorphic function of the complex coordinates
$Z^{i}$.

An object $X$ is said to have K\"ahler weight $q$ when $X$ transforms under
the above K\"ahler transformations as

\begin{equation}
    X\rightarrow e^{-(q\lambda-q\lambda^{*})/2}X\,.
\end{equation}

\noindent
The K\"ahler-covariant derivative $\mathfrak{D}$ acting
on X has the following holomorphic and anti-holomorphic components

\begin{equation}
\label{eq:Kcovariantderivative} \mathfrak{D}_{i}X \equiv
\left(\nabla_{i} +iq \mathcal{Q}_{i}\right)X\, , \hspace{1cm}
\mathfrak{D}_{i^{*}}X \equiv\left( \nabla_{i^{*}} -i\bar{q}
\mathcal{Q}_{i^{*}}\right)X\, ,
\end{equation}

\noindent
where $\nabla$ is the standard covariant derivative associated to the
Levi-Civit\`a connection, Eqs.~\eqref{eq:KCCChrisSymb}, on $\mathcal{M}$. For
objects with K\"ahler weight $q$ the space-time pullback of the
K\"ahler-covariant derivative is given by

\begin{equation}
\label{eq:Kcovariantderivative2}
\mathfrak{D}_{\mu}= \nabla_{\mu} +iq\mathcal{Q}_{\mu}\, ,
\end{equation}

\noindent
where $\nabla_{\mu}$ is the standard space-time covariant derivative plus the
pullback of the Levi-Civit\`a connection on $\mathcal{M}$ if necessary and
where $\mathcal{Q}_{\mu}$ is the pullback of the K\"ahler 1-form of
Eq.~\eqref{eq:K1form}.

A special K\"{a}hler manifold is the base manifold of a
$Sp(2n_V+2,\mathbb{R})\times U(1)$ bundle \cite{kn:toinereview}.  There exist
sections $\mathcal{V}$ such that

\begin{equation}
\label{eq:SGDefFund}
\mathcal{V} =
\left(
\begin{array}{c}
\mathcal{L}^{\Lambda}\\
\mathcal{M}_{\Sigma}\\
\end{array}
\right) \;\; \rightarrow \;\;
\left\{
\begin{array}{lcl}
\langle \mathcal{V}\mid\mathcal{V}^{*}\rangle
& \equiv &
\mathcal{L}^{*\, \Lambda}\mathcal{M}_{\Lambda}
-\mathcal{L}^{\Lambda}\mathcal{M}^{*}_{\Lambda}
= -i\, , \\
& & \\
\mathfrak{D}_{i^{*}}\mathcal{V} & = & (\partial_{i^{*}}-
{\textstyle\frac{1}{2}}\partial_{i^{*}}\mathcal{K})\mathcal{V} =0 \, ,\\
& & \\
\langle\mathfrak{D}_{i}\mathcal{V}\mid\mathcal{V}\rangle & = & 0 \, ,
\end{array}
\right.
\end{equation}

\noindent
where $\mathfrak{D}_{i}\mathcal{V}=(\partial_{i}+
{\textstyle\frac{1}{2}}\partial_{i}\mathcal{K})\mathcal{V}$.

It follows from the basic definitions, Eqs.~\eqref{eq:SGDefFund},
that

\begin{equation}
\label{eq:SGProp1}
\begin{array}{rclrcl}
\mathfrak{D}_{i^{*}}\ \mathfrak{D}_{i}\mathcal{V} & = &
\mathcal{G}_{ii^{*}} \ \mathcal{V}\,,\hspace{2cm} &
\langle\mathfrak{D}_{i}\mathcal{V}\mid\mathfrak{D}_{i^*}\mathcal{V}^*\rangle
& = &
i\,\mathcal{G}_{ii^{*}} \, , \\
& & & & & \\
\langle\mathfrak{D}_{i}\mathcal{V}\mid\mathcal{V}^{*}\rangle &  = & 0\, , &
\langle\mathfrak{D}_{i}\mathcal{V}\mid\mathcal{V}\rangle & = & 0 \, ,\\
& & & & & \\
\langle\mathfrak{D}_{i}\mathfrak{D}_{j}\mathcal{V}\mid\mathcal{V}\rangle
& = & 0\,, & \langle \mathfrak{D}_{j}\mathcal{V}\mid
\mathfrak{D}_{i}\mathcal{V}\rangle & = & 0\, .
\end{array}
\end{equation}

\noindent
If we now group together $\mathcal{V}$ and $\mathfrak{D}_{i}\mathcal{V}$ into
$\mathcal{E}_{\Lambda} = (\mathcal{V},\mathfrak{D}_{i}\mathcal{V})$ we can see
that $\langle \mathcal{E}_{\Sigma}\mid\mathcal{E}^{*}{}_{\Lambda}\rangle$ is a
non-degenerate matrix. Using
$\{\mathcal{E}_{\Sigma},\mathcal{E}^{*}{}_{\Lambda}\}$ as a basis for the
space of symplectic sections we obtain the following completeness relation

\begin{equation}
\label{completeness}
i\mathbbm{1} =-\mid\mathcal{V}^{*}\rangle\langle\mathcal{V}\mid
+\mid\mathcal{V}\rangle\langle\mathcal{V}^{*}\mid
-\mathcal{G}^{ii^{*}}\mid\mathfrak{D}_{i}\mathcal{V}\rangle\langle\mathfrak{D}_{i^*}\mathcal{V}^*\mid
+\mathcal{G}^{ii^{*}}
\mid\mathfrak{D}_{i^*}\mathcal{V}^*\rangle\langle\mathfrak{D}_{i}\mathcal{V}\mid\, .
\end{equation}

We write for the components of $\mathfrak{D}_{i}\mathcal{V}$ the following

\begin{equation}
\label{eq:SGDefU}
\mathfrak{D}_{i}\mathcal{V}=\left(
\begin{array}{c}
f^{\Lambda}{}_{i}\\
h_{\scriptscriptstyle{\Sigma}\, i}
\end{array}
\right)\, .
\end{equation}

\noindent
The period matrix $\mathcal{N}_{\Lambda\Sigma}$ is defined
by the following two relations

\begin{equation}
\label{eq:SGDefN}
\mathcal{M}_{\Lambda} = \mathcal{N}_{\Lambda\Sigma} \mathcal{L}^{\Sigma}\, ,
\hspace{1cm}
h_{\Lambda\, i} = \mathcal{N}^{*}{}_{\Lambda\Sigma} f^{\Sigma}{}_{i} \, .
\end{equation}

\noindent
The identity $\langle\mathfrak{D}_{i}\mathcal{V}\mid\mathcal{V}^*\rangle =0$
implies that $\mathcal{N}$ is symmetric in its symplectic indices.

From the properties, Eqs.~\eqref{eq:SGDefFund}, one concludes that
$\mathcal{V}$ transforms under K\"{a}hler transformations as

\begin{equation}\label{KahlertrafoV}
    \mathcal{V}\rightarrow e^{-\tfrac{1}{2}(\lambda-\lambda^{*})}\mathcal{V}\,.
\end{equation}

For further details and identities the interested reader can consult the basic
references
\cite{Andrianopoli:1996cm,Ceresole:1995ca,Ceresole:1995jg,Craps:1997gp}, the
review \cite{kn:toinereview} or Ref.~\cite{Meessen:2006tu,Bellorin:2005zc}
whose conventions and results we follow.

\section{Quaternionic K\"ahler geometry}
\label{sec-QKG}

A quaternionic K\"ahler manifold is a real $4n_H$-dimensional
Riemannian manifold $\mathsf{HM}$ endowed with a triplet of
complex structures $\mathsf{J}^{x}: T(\mathsf{HM})\rightarrow
T(\mathsf{HM})\, ,\,\,\, (x=1,2,3)$ that satisfy the quaternionic
algebra

\begin{equation}
\mathsf{J}^{x}  \mathsf{J}^{y} = -\delta^{xy}
+\varepsilon^{xyz}\mathsf{J}^{z}\, ,
\end{equation}

\noindent
and with respect to which the metric, denoted by $\mathsf{H}$, is Hermitean

\begin{equation}
\mathsf{H}(\ \mathsf{J}^{x} X,\ \mathsf{J}^{x}Y\ )= \mathsf{H}(X,Y)\, ,
\hspace{1cm}
\forall X,Y \in  T(\mathsf{HM})\, .
\end{equation}

\noindent This implies the existence of a triplet of 2-forms
$\mathsf{K}^{x}(X,Y)\equiv \mathsf{H}(\ \mathsf{J}^{x}X ,Y)$
globally known as the $\mathfrak{su}(2)$-valued hyperK\"ahler
2-forms.

The structure of a quaternionic K\"ahler manifold requires an $SU(2)$
bundle to be constructed over $\mathsf{HM}$ with connection 1-form
$\mathsf{A}^{x}$ with respect to which the hyperK\"ahler 2-form is covariantly
closed, i.e.

\begin{equation}
\mathfrak{D}\mathsf{K}^{x}\; \equiv\; d\mathsf{K}^{x}
+\varepsilon^{xyz}\ \mathsf{A}^{y}\wedge \mathsf{K}^{z}\; =\; 0\, .
\end{equation}

\noindent Then if the curvature of this bundle

\begin{equation}
\mathsf{F}^{x}\; \equiv\; d\mathsf{A}^{x}
\ +\ {\textstyle\frac{1}{2}}\varepsilon^{xyz}\
     \mathsf{A}^{y} \wedge \mathsf{A}^{z}\, ,
\end{equation}

\noindent is equal to minus the hyperK\"ahler 2-form

\begin{equation}
\mathsf{F}^{x}=-\mathsf{K}^{x}\, ,
\end{equation}

\noindent the manifold is a quaternionic K\"ahler manifold as it
appears in supergravity.

The $SU(2)$ connection acts on objects with vectorial $SU(2)$ indices,
such as the chiral spinors in this article, as follows

\begin{eqnarray}
\mathfrak{D} \xi_{I} & \equiv  & d\xi_{I} +\mathsf{A}_{I}{}^{J}\xi_{J}\, ,\label{DxiIdown}\\
& & \nonumber\\
\mathfrak{D} \chi^{I} & \equiv  & d\chi^{I} +\mathsf{A}^{I}{}_{J}\chi^{J}\, .\label{DxiIup}
\end{eqnarray}

\noindent
The vector $SU(2)$ indices on $\mathsf{A}^{I}{}_{J}$ are raised and lowered under complex conjugation as

\begin{equation}
\mathsf{A}^{I}{}_{J} = (\mathsf{A}_{I}{}^{J})^{*}\, .
\end{equation}

\noindent Following Ref.~\cite{Andrianopoli:1996cm} we put

\begin{equation}
\mathsf{A}_{I}{}^{J}\; \equiv\;
{\textstyle\frac{i}{2}}\ \mathsf{A}^{x}\ (\sigma_{x})_{I}{}^{J}\,,
\end{equation}

\noindent and similarly for the curvature $\mathsf{F}_{I}{}^{J}$
where the 3 matrices $(\sigma_{x})_{I}{}^{J}$ are the Pauli
matrices.

The holonomy group of a quaternionic K\"{a}hler manifold
$\mathsf{HM}$ is $Sp(1)\times Sp(2n_H)$ where $Sp(2n_H)\simeq
U(4n_H)\cap Sp(4n_H,\mathbb{C})$, so that $Sp(1)\simeq SU(2)$. It
is convenient to use a Vielbein on $\mathsf{HM}$, denoted by

\begin{equation}
\mathsf{U}^{\alpha I} \; =\; \mathsf{U}^{\alpha I}{}_{u}\ dq^{u}\, ,\hspace{.5cm}
\text{where}\hspace{.5cm} u\ =\ 1,\ldots ,4n_H\,,
\end{equation}

\noindent
having as `flat' indices a pair $\alpha I$ consisting of one $Sp(2n_H)$ index $\alpha = 1,\ldots,2n_H$
and one $SU(2)$ index $I=1,2$. We shall refer to this object as the
Quadbein. This Quadbein is related to the metric
$\mathsf{H}_{uv}$ by

\begin{equation}
\mathsf{H}_{uv} \; =\; \mathsf{U}^{\alpha I}{}_{u}\ \mathsf{U}^{\beta J}{}_{v}\
\varepsilon_{IJ}\mathbb{C}_{\alpha\beta}\, ,
\end{equation}

\noindent where $\varepsilon_{IJ}=-\varepsilon_{JI}$ and
$\mathbb{C}_{\alpha\beta}=-\mathbb{C}_{\beta\alpha}$ are the flat
$Sp(2n_H)$ and $SU(2)$ invariant metrics. It is required that

\begin{equation}
  \begin{array}{rcl}
2\ \mathsf{U}^{\alpha I}{}_{(u}\ \mathsf{U}^{\beta J}{}_{v)}\
\mathbb{C}_{\alpha\beta} & = & \mathsf{H}_{uv} \varepsilon^{IJ}\, ,\\
& & \\
\mathsf{U}_{\alpha I\, u} & \equiv & (\mathsf{U}^{\alpha I}{}_{u})^{*}
\; =\; \varepsilon_{IJ}\mathbb{C}_{\alpha\beta}\ \mathsf{U}^{\beta J}{}_{u}\, .
\end{array}
\end{equation}

\noindent
The inverse Quadbein $\mathsf{U}^{u}{}_{\alpha I}$ satisfies

\begin{equation}
\mathsf{U}_{\alpha I}{}^{u}\ \mathsf{U}^{\alpha I}{}_{v}
=\delta^{u}{}_{v}\, .
\end{equation}

%

For further details and identities see e.g.~Refs.~\cite{Andrianopoli:1996cm,Bergshoeff:2002qk,Bergshoeff:2004nf},
the review \cite{kn:toinereview} or Ref.~\cite{Huebscher:2006mr} whose
conventions and results we follow and use.



\begin{thebibliography}{99}

\bibitem{Cremmer:1998px}
E.~Cremmer, B.~Julia, H.~Lu and C.~N.~Pope,
Nucl.\ Phys.\  B {\bf 535} (1998) 242
[\hepth{9806106}].

\bibitem{Bergshoeff:2007aa}
E.~Bergshoeff, J.~Hartong and D.~Sorokin,
\arxiv{0708.2287} [hep-th].


\bibitem{Meessen:1998qm}
P.~Meessen and T.~Ort\'{\i}n,
Nucl.\ Phys.\  B {\bf 541} (1999) 195
[\hepth{9806120}].

\bibitem{Dall'Agata:1998va}
G.~Dall'Agata, K.~Lechner and M.~Tonin,
JHEP {\bf 9807} (1998) 017
[\hepth{9806140}].



\bibitem{Bergshoeff:2006jj}
E.~A.~Bergshoeff, J.~Hartong, T.~Ort\'{\i}n and D.~Roest,
JHEP {\bf 0702} (2007) 003
[\hepth{0612072}].

\bibitem{Greene:1989ya}
B.~R.~Greene, A.~D.~Shapere, C.~Vafa and S.~T.~Yau,
Nucl.\ Phys.\  B {\bf 337} (1990) 1.

\bibitem{Freedman:1980us}
D.~Z.~Freedman and P.~K.~Townsend,
Nucl.\ Phys.\  B {\bf 177} (1981) 282.

\bibitem{Meessen:2006tu}
P.~Meessen and T.~Ort\'{\i}n,
Nucl.\ Phys.\  B {\bf 749} (2006) 291
[\hepth{0603099}].

\bibitem{Bellorin:2005zc}
J.~Bellor\'{\i}n and T.~Ort\'{\i}n,
Nucl.\ Phys.\  B {\bf 726} (2005) 171
[\hepth{0506056}].

\bibitem{Andrianopoli:1996cm}
L. Andrianopoli, M. Bertolini, A. Ceresole, R. D'Auria, S. Ferrara,
P. Fr\'e and T. Magri,
J.\ Geom.\ Phys.\  {\bf 23} (1997) 111
[\hepth{9605032}].

\bibitem{kn:toinereview}
A. van Proeyen,
lectures given at the Institute Henri Poincar\'e, Paris, November 2000.
\href{http://itf.fys.kuleuven.ac.be/~toine/LectParis.pdf}{\tt http://itf.fys.kuleuven.ac.be/\~{}toine/LectParis.pdf}

\bibitem{deWit:1984pk}
B. de Wit and A. van Proeyen,
Nucl.\ Phys.\ B {\bf 245} (1984) 89.

\bibitem{deWit:1984px}
B. de Wit, P.G. Lauwers and A. van Proeyen,
Nucl.\ Phys.\ B {\bf 255} (1985) 569.

\bibitem{Gaillard:1981rj}
M.~K.~Gaillard and B.~Zumino,
Nucl.\ Phys.\ B {\bf 193} (1981) 221.

\bibitem{Bergshoeff:2006gs}
E.~A.~Bergshoeff, M.~de Roo, S.~F.~Kerstan, T.~Ort\'{\i}n and F.~Riccioni,
JHEP {\bf 0702} (2007) 007
[\hepth{0611036}].

\bibitem{Behrndt:1997ny}
K.~Behrndt, D.~L\"ust and W.~A.~Sabra,
Nucl.\ Phys.\ B {\bf 510} (1998) 264
[\hepth{9705169}].

\bibitem{LopesCardoso:2000qm}
G.~Lopes Cardoso, B.~de Wit, J.~Kappeli and T.~Mohaupt,
JHEP {\bf 0012} (2000) 019
[\hepth{0009234}].

\bibitem{Bellorin:2006xr}
J.~Bellor\'{\i}n, P.~Meessen and T.~Ort\'{\i}n,
Nucl.\ Phys.\  B {\bf 762} (2007) 229
[\hepth{0606201}].


\bibitem{Claus:1997fk}
 P.~Claus, B.~de Wit, M.~Faux, B.~Kleijn, R.~Siebelink and P.~Termonia,
   Nucl.\ Phys.\ B {\bf 512} (1998) 148
 [\hepth{9710212}].


\bibitem{deAzcarraga:1989gm}
J.~A.~de Azcarraga, J.~P.~Gauntlett, J.~M.~Izquierdo and P.~K.~Townsend,
Phys.\ Rev.\ Lett.\  {\bf 63} (1989) 2443.

\bibitem{deWit:1995tf}
B.~de Wit and A.~Van Proeyen,
[\hepth{9505097}].

\bibitem{Atick:1985iy}
J.~J.~Atick, A.~Dhar and B.~Ratra,
Phys.\ Lett.\  B {\bf 169} (1986) 54.

\bibitem{Howe:1995zm}
P.~S.~Howe, J.~M.~Izquierdo, G.~Papadopoulos and P.~K.~Townsend,
Nucl.\ Phys.\ B {\bf 467}, 183 (1996)
[\hepth{9505032}].


\bibitem{Theis:2003jj}
 U.~Theis and S.~Vandoren,
 JHEP {\bf 0304} (2003) 042
 [\hepth{0303048}].


\bibitem{Aoyama:2005hb}
  S.~Aoyama,
  Phys.\ Lett.\  B {\bf 625} (2005) 127
  [\hepth{0506248}].

\bibitem{Huebscher:2006mr}
M.~H\"ubscher, P.~Meessen and T.~Ort\'{\i}n,
Nucl.\ Phys.\ B {\bf 759} (2006) 228
[\hepth{0606281}].


\bibitem{Bergshoeff:1998re}
E.~Bergshoeff, E.~Eyras, R.~Halbersma, J.~P.~van der Schaar, C.~M.~Hull and Y.~Lozano,
Nucl.\ Phys.\  B {\bf 564} (2000) 29
[\hepth{9812224}].

\bibitem{Riccioni:2004nr}
F.~Riccioni,
Nucl.\ Phys.\  B {\bf 711} (2005) 231
[\hepth{0410185}].

\bibitem{Bergshoeff:2005ac}
E.~A.~Bergshoeff, M.~de Roo, S.~F.~Kerstan and F.~Riccioni,
JHEP {\bf 0508} (2005) 098
[\hepth{0506013}].

\bibitem{Bergshoeff:2006qw}
E.~A.~Bergshoeff, M.~de Roo, S.~F.~Kerstan, T.~Ortin and F.~Riccioni,
JHEP {\bf 0607} (2006) 018
[\hepth{0602280}].

\bibitem{Riccioni:2007au}
  F.~Riccioni and P.~West,
  JHEP {\bf 0707} (2007) 063
  [\arxiv{0705.0752} [hep-th]].

\bibitem{Bergshoeff:2007qi}
  E.~A.~Bergshoeff, I.~De Baetselier and T.~A.~Nutma,
  JHEP {\bf 0709}, 047 (2007)
  [\arxiv{0705.1304} [hep-th]].

\bibitem{Bergshoeff:2007vb}
  E.~A.~Bergshoeff, J.~Gomis, T.~A.~Nutma and D.~Roest,
  \arxiv{0711.2035} [hep-th].

\bibitem{Gomis:2007gb}
  J.~Gomis and D.~Roest,
  \arxiv{0706.0667} [hep-th].




\bibitem{Bergshoeff:1997mg}
E.~Bergshoeff, M.~de Roo and E.~Eyras,
Phys.\ Lett.\  B {\bf 413} (1997) 70
[\hepth{9707130}].

\bibitem{Lavrinenko:1997qa}
I.~V.~Lavrinenko, H.~Lu and C.~N.~Pope,
Class.\ Quant.\ Grav.\  {\bf 15} (1998) 2239
[\hepth{9710243}].

\bibitem{Kaloper:1998kr}
N.~Kaloper, R.~R.~Khuri and R.~C.~Myers,
Phys.\ Lett.\  B {\bf 428} (1998) 297
[\hepth{9803066}].


\bibitem{Hull:1998vy}
C.~M.~Hull,
JHEP {\bf 9811} (1998) 027
[\hepth{9811021}].

\bibitem{Cowdall:2000sq}
P.~M.~Cowdall,
\hepth{0009016}.

\bibitem{Hull:2002wg}
C.~M.~Hull,
Class.\ Quant.\ Grav.\  {\bf 21} (2004) 509
[\hepth{0203146}].

\bibitem{Bergshoeff:2002mb}
E.~Bergshoeff, U.~Gran and D.~Roest,
Class.\ Quant.\ Grav.\  {\bf 19} (2002) 4207
[\hepth{0203202}].

\bibitem{Bergshoeff:1996ui}
E.~Bergshoeff, M.~de Roo, M.~B.~Green, G.~Papadopoulos and P.~K.~Townsend,
Nucl.\ Phys.\  B {\bf 470} (1996) 113
[\hepth{9601150}].

\bibitem{Ceresole:1995ca}
A. Ceresole, R. D'Auria and S. Ferrara,
Nucl.\ Phys.\ Proc.\ Suppl.\  {\bf 46} (1996) 67
[\hepth{9509160}].

\bibitem{Ceresole:1995jg}
A. Ceresole, R. D'Auria, S. Ferrara and A. van Proeyen,
Nucl.\ Phys.\ B {\bf 444} (1995) 92
[\hepth{9502072}].

\bibitem{Craps:1997gp}
B. Craps, F. Roose, W. Troost and A. van Proeyen,
Nucl.\ Phys.\ B {\bf 503} (1997) 565
[\hepth{9703082}].



\bibitem{Bergshoeff:2002qk}
 E.~Bergshoeff, S.~Cucu, T.~De Wit,  J.~Gheerardyn, R.~Halbersma, S.~Vandoren and A.~Van Proeyen,
 JHEP {\bf 0210} (2002) 045
 [\hepth{0205230}].

\bibitem{Bergshoeff:2004nf}
 E.~Bergshoeff, S.~Cucu, T.~de Wit, J.~Gheerardyn,  S.~Vandoren and A.~Van Proeyen,
 Commun.\ Math.\ Phys.\ {\bf 262} (2006) 411
 [\hepth{0411209}].
















\end{thebibliography}
\end{document}